\documentclass[aps,pra,10pt,twocolumn,superscriptaddress,floatfix,nofootinbib,showpacs,longbibliography]{revtex4-2}
\usepackage[utf8]{inputenc}  
\usepackage{amsfonts}
\usepackage[table]{xcolor}
\usepackage[british]{babel}  
\usepackage[sc,osf]{mathpazo}
\usepackage[scaled=0.86]{berasans}  
\usepackage[colorlinks=true, citecolor=blue, urlcolor=blue]{hyperref}  
\usepackage{graphicx} 
\usepackage[babel]{microtype}  
\usepackage{amsmath,amssymb,amsthm,bm,amsfonts,mathrsfs,bbm} 

\usepackage{xspace}  
\usepackage{pgf,tikz}
\usepackage{xcolor}
\usepackage{multirow}
\usepackage{array}
\usepackage{bigstrut}
\usepackage{braket}
\usepackage{color}
\usepackage{natbib}
\usepackage{multirow}
\usepackage{mathtools}
\usepackage{float}
\usepackage[caption = false]{subfig}
\usepackage{xcolor,colortbl}
\usepackage{color}
\usepackage{rotating}
\usepackage{tikz}

\usepackage{listings}
\usepackage{braket}
\usepackage{play}
\usepackage{epsfig}
\usepackage{soul}

\newcommand{\be}{\begin{equation}}
\newcommand{\ee}{\end{equation}}
\newcommand{\ba}{\begin{eqnarray}}
\newcommand{\ea}{\end{eqnarray}}

\newcommand{\tr}{\operatorname{Tr}}

\usepackage{centernot}
\usepackage{subfig}

\definecolor{codegreen}{rgb}{0,0.6,0}
\definecolor{codegray}{rgb}{0.5,0.5,0.5}
\definecolor{codepurple}{rgb}{0.58,0,0.82}
\definecolor{backcolour}{rgb}{0.95,0.95,0.92}

\lstdefinestyle{mystyle}{
    backgroundcolor=\color{backcolour},   
    commentstyle=\color{codegreen},
    keywordstyle=\color{magenta},
    numberstyle=\tiny\color{codegray},
    stringstyle=\color{codepurple},
    basicstyle=\ttfamily\footnotesize,
    breakatwhitespace=false,         
    breaklines=true,                 
    captionpos=b,                    
    keepspaces=true,                 
    numbers=left,                    
    numbersep=5pt,                  
    showspaces=false,                
    showstringspaces=false,
    showtabs=false,                  
    tabsize=2
}

\begin{document}
	
\title{Exploring super-additivity of coherent information of noisy quantum channels through Genetic algorithms}

\author{Govind Lal Sidhardh}
\affiliation{School of Physics, IISER Thiruvananthapuram, Vithura, Kerala 695551, India.}

\author{Mir Alimuddin}
\affiliation{School of Physics, IISER Thiruvananthapuram, Vithura, Kerala 695551, India.}
\affiliation{Department of Theoretical Sciences, S. N. Bose National Center for Basic Sciences, Block JD, Sector III, Salt Lake, Kolkata 700106, India.}

\author{Manik Banik}
\affiliation{Department of Theoretical Sciences, S. N. Bose National Center for Basic Sciences, Block JD, Sector III, Salt Lake, Kolkata 700106, India.}

\begin{abstract}
Machine learning techniques are increasingly being used in fundamental research to solve various challenging problems. Here we explore one such technique to address an important problem in quantum communication scenario. While transferring quantum information through a noisy quantum channel, the utility of the channel is characterized by its quantum capacity. Quantum channels, however, display an intriguing property called super-additivity of coherent information. This makes the calculation of quantum capacity a hard computational problem involving optimization over an exponentially increasing search space. In this work, we first utilize a neural network ansatz to represent quantum states and then apply an evolutionary optimization scheme to address this problem. We find regions in the three-parameter space of qubit Pauli channels where coherent information exhibits this super-additivity feature. We characterised the quantum codes that achieves high coherent information, finding several non-trivial quantum codes that outperforms the repetition codes for some Pauli channels. For some Pauli channels, these codes displays very high super-additivity of the order of 0.01, much higher than the observed values in other well studied quantum channels. We further compared the learning performance of the Neural Network ansatz with the raw ansatz to find that in the three-shot case, the neural network ansatz outperforms the raw representation in finding quantum codes of high coherent information. We also compared the learning performance of the evolutionary algorithm with a simple Particle Swarm Optimisation scheme and show empirical results indicating comparable performance, suggesting that the Neural Network ansatz coupled with the evolutionary scheme is indeed a promising approach to finding non-trivial quantum codes of high coherent information.
\end{abstract}

\maketitle

\section{Introduction}
The primary goal of any communication protocol is the reliable transfer of information from one spacetime point to another. However, almost in all practical scenarios, the communication lines get interrupted with unavoidable noises, and hence their successful implementation crucially depends on mitigation of those undesirable noises. From a rigorous mathematical point of view, the problem of communication in the presence of noise was first addressed by Claude Shannon in his seminal $1948$ paper \cite{Shannon1948}. Shannon modeled a noisy classical communication line, also called a classical channel, as a stochastic map. This map is from some random input variable used by a sender for encoding information to some output random variable received by a distant receiver. The utility of such a classical channel is then characterized by its capacity, which is given by the mutual information between input and output random variables - an entropic quantity - optimized over all possible input probability distributions. Interestingly, the additivity property of mutual information makes parallel uses of many identical copies of a channel equivalent to multiple uses of its single-copy \cite{Cover2005}. Furthermore, the devices used for storing and transferring information are classical objects whose working rules are fundamentally governed by classical physics.  

Quantum information theory utilizes peculiar features of the quantum world to devise novel communication protocols that are advantageous over their classical counterpart \cite{Nielsen2000}, and in some cases do not have any classical analog \cite{Bennett92,Bennett93,Bennett2014,PRXQuantum.2.020350,Chiribella_2021,Guha2021quantumadvantage,https://doi.org/10.1002/andp.202000334,PhysRevA.104.012420,agrawal2021better}. While modeling the quantum communication scenario, classical input and output variables are replaced by respective input and output Hilbert spaces, and accordingly, the channel action is most generally described by a completely positive trace preserving (CPTP) map from the operator space on input to operator space on output Hilbert space \cite{wilde2011classical}. Depending on what purpose a quantum channel is used for, its merit of utility is quantified by different quantities. For instance, the classical capacity of a quantum channel quantifies how good it is to transfer classical information \cite{Schumacher1997,Holevo1998}. In contrast, the private capacity captures its efficacy to transfer classical information in a secure way from a sender to a receiver so that no eavesdropper can know it \cite{Devetak2005}. On the other hand, the quantum capacity of a quantum channel captures its utility to transfer quantum information \cite{Lloyd1997,Shor2002,Devetak2005}. While the quantum capacity is defined through an entropic quantity -- coherent information of its multiple uses in the asymptotic limit -- obtaining a single-letter expression for this quantity is extremely difficult in general. Unlike the classical capacity of a classical channel, the quantum capacity of a quantum channel can exhibit a striking super-additive phenomenon\footnote{Importantly, classical capacity and private capacity of quantum channels also exhibit super-additive behavior \cite{Hastings2009,Li2009,Smith2009}.}, where two different channels each with zero quantum capacity can be used together to send quantum information with nonzero rate \cite{Smith2008,Oppenheim2008,Smith2011}. Later, it has been shown that there exists a quantum channel whose $N$-parallel use is no good to send quantum information whereas its $(N+1)$-parallel use becomes useful for transferring quantum information, with $N$ taking arbitrary integer values \cite{Cubitt2015}. The super-additivity phenomenon makes the study of quantum communication more interesting as parallel use of many copies of a channel can be more beneficial than multiple uses of a single channel. This phenomenon of super-additivity has been explicitly demonstrated in the case of the one-parameter family of Depolarising channels, for which the repetition codes are known to display super-additivity at three channel-use and five channel-use level \cite{Bausch2020, dep1, dep2, dep3}. More recently, super-additivity of coherent information was demonstrated for two channel-uses for the two-parameter family of Dephrasure Channels again using the repetition code \cite{Leditzky2018,Yu2020}. Recently, quantum codes based on w-states were shown to display super-additivity for generalized erasure channels \cite{Filippov_2021m}. However, for more general classes of quantum channels, such simple quantum codes may not display the super-additivity feature, and therefore, one needs to enlarge the search over the code space.
\par
In this work, we utilize a machine learning-based optimization technique, as pioneered in Ref.\cite{Bausch2020}, to explore super-additivity of coherent information for a general class of qubit Pauli channels that includes Depolarising Channel as a special case. This study of Pauli channels is important because of its considerable generality and the practical utility of modelling more general quantum noises as Pauli channels using techniques like Pauli projection \cite{newcite3} in quantum error correction applications\cite{newcite4}. For the class of qubit Depolarising channels, no two-qubit state is known to display super-additivity of coherent information \cite{Leditzky2018}. Does this carry over to the class of generalized Pauli channels as well? If not, in what regions of the three-parameter space can this super-additivity be proven. In this work, we attempt to address these questions by building on the ideas from Ref.\cite{Bausch2020}. Specifically, we use genetic algorithms on the neural network ansatz to map out regions of super-additivity and characterize quantum codes of high coherent information for the three-parameter family of qubit Pauli channels. In addition, we also investigate the learning performance of the scheme by comparing it with cases where (1) a raw ansatz is used instead of the Neural Network ansatz; and (2) other cases where Particle Swarm Optimisation is used instead of GAs.
 \par
The remaining sections of the paper are organized
along these lines. In Section \ref{secIIA}, a brief introduction to the problem of Quantum capacities is given. A brief introduction to genetic algorithms and methods are given in Section \ref{secIIC} and Section \ref{secIID}, respectively. The main findings in our work have been presented in Section \ref{secIV}. Lastly, we put our conclusions in Section \ref{secV}. Detailed analysis of numerical learning performance is given in the Appendix \ref{Appendix}.

\section{preliminaries}
In this section, we first review the idea of super-additivity of coherent information of a quantum channel, and then we will briefly discuss some concepts in machine learning that will be relevant to our purpose. 

\subsection{Super-additivity of coherent information and quantum capacity}\label{secIIA}
Before moving to the quantum case, let us recall that a noisy classical channel is a stochastic map $\mathcal{N}:P(X)\to P(Y)$, where $P(X)$ and $P(Y)$ respectively denote probability distributions on the input and output random variables $X~\&~Y$.  Shannon's Noisy channel coding theorem \cite{Shannon1948} specifies capacity of such a channel as
\begin{align}
\mathcal{C}(\mathcal{N})=\max_{P(X)}H(X:Y)\label{C(N)},  
\end{align}            
where, $H(X:Y):=H(X)+H(Y)-H(X Y)$ is the mutual information between the input and output distributions with $H(X):=-\sum_{x\in X}p(x)\log p(x)$ being the Shannon entropy, and the optimization is over the input probability distributions $P(X)\equiv\{p(x)~|~p(x)\ge0~\&~\sum_{x\in X}p(x)=1\}$.

In quantum scenario, a communication process is more generally described by a channel (CPTP map) $\Lambda:\mathcal{L}(\mathcal{H}_{in})\to\mathcal{L}(\mathcal{H}_{out})$, where $\mathcal{L}(\mathcal{Z})$ is the operator space acting on the Hilbert space $\mathcal{Z}$ and $\mathcal{H}_{in},~\mathcal{H}_{out}$ are the respective Hilbert spaces for the input and output systems of the channel. Complete positivity of $\Lambda$ assures positivity of the map when it is applied on the part of a composite systems, {\it i.e.} $(\Lambda_A\otimes\mathbb{I}_R)\rho_{AR}\ge0~\forall~\rho_{AR}\in\mathcal{P}(\mathcal{H}_{A}\otimes\mathcal{H}_R)$, where $\mathcal{P}(\mathcal{Z})\subset\mathcal{L}(\mathcal{Z})$ denotes the set of positive operators acting on $\mathcal{Z}$ and $\mathbb{I}_{z}$ denotes the identity map on $\mathcal{L}(\mathcal{Z})$. Furthermore, trace preserving condition implies $\tr[\Lambda_z(\rho)]=1~\forall~\rho\in\mathcal{D}(\mathcal{Z})\subset\mathcal{P}(\mathcal{Z})$, where $\mathcal{D}(\mathcal{Z})$ denotes the set of density operators acting on $\mathcal{Z}$. The quantum capacity $Q(\Lambda)$ of a quantum channel $\Lambda$ is given by the following regularized expression \cite{Lloyd1997,Shor2002,Devetak2005}:
\begin{align}
Q(\Lambda)&:=\lim_{n\rightarrow \infty} \frac{1}{n}Q^{(1)}(\Lambda^{\otimes n})\label{Q(N)},\\\nonumber 
&\mbox{where},\\
Q^{(1)}(\Lambda^{\otimes n})&:=\max_{\rho_S} \mathcal{I}(\rho_S,\Lambda^{\otimes n}). \label{CI}
\end{align}
is the $n$-shot channel capacity, where $n$ identical copies of $\Lambda$ are used in parallel. The maximization in Eq.(\ref{CI}) is over all valid quantum states $\rho_S\in\mathcal{D}(\mathcal{H}_{in}^{\otimes n})$. In general $\rho_S$'s  are mixed states, although in calculations it is often convenient to introduce a purified state $\ket{\Psi_{SR}}\in \mathbb{C}^{2n}$ such that $Tr_R(\ket{\Psi_{SR}}\bra{\Psi_{SR}})=\rho_S$. With this purified state, expression for the quantity $\mathcal{I}$, called coherent information of $\Lambda^{\otimes n}$ for the input $\rho_S$, reads as
\begin{align}
\mathcal{I}\left(\rho_S,\Lambda^{\otimes n}\right)&\equiv\mathcal{I}\left(\ket{\psi_{SR}},\Lambda^{\otimes n}\right)\nonumber\\
&= S\left(\Lambda^{\otimes n}\left(\rho_S\right)\right)-S\left(\Lambda ^{\otimes n}\otimes \mathbb{I}_R\left(\ket{\psi_{SR}}\bra{\psi_{SR}}\right)\right).
\end{align} 
Here $S(\rho)=-\tr(\rho\log(\rho))$ is the Von-Neumann entropy\cite{Nielsen2000}, and $\mathbb{I}_R$ is the identity channel on the purifying environment $R$. Coherent information of a channel $\Lambda$ is then defined as $\mathcal{I}\left(\Lambda\right)\equiv\max_{\ket{\psi_{SR}}}\mathcal{I}\left(\ket{\psi_{SR}},\Lambda\right)$.

Note that, the expression of capacity in Eq.(\ref{C(N)}) for a classical channel follows from a similar regularized formula by invoking the additivity of mutual information, {\it i.e.} $\mathcal{C}(\mathcal{N}^{\otimes n})=n \mathcal{C}(\mathcal{N})$ \cite{Cover2005}. This greatly simplifies the optimization problem over an unbounded number channel uses to a tractable one in the single-letter space. One might hope that such a simplification might exist for the quantum scenario as well; however, that is not the case! There are a few class of channels that exhibits weak additivity, like the degradable channel, for which the quantum capacity reduces to coherent information \cite{Devetak2005}, but this is not true in general. In fact, one can find channels that have strictly super-additive coherent information, {\it i.e.} $\mathcal{I}(\mathcal{N}^{\otimes n})>n\mathcal{I}(\mathcal{N})$ \cite{Smith2008,Oppenheim2008,Smith2011,Cubitt2015,Bausch2020,newcite5}. This is quite interesting as parallel use of multi-copy channels with properly constructed quantum codes can yield more transmission rates than using the channel multiple times. Unfortunately, this appears with a downside; the optimization needs to be carried out over an unbounded number of channel uses, rendering the problem intractable. Yet all is not lost as it is known that $\frac{1}{n}Q^{(1)}(\Lambda^{\otimes n})$ is lower bound to the quantum capacity \cite{Lloyd1997,Shor2002,Devetak2005}, {\it i.e.}
\begin{align}
   Q(\Lambda)\geq \frac{1}{n}Q^{(1)}\left(\Lambda^{\otimes n}\right)\geq \frac{1}{n} \mathcal{I}\left(\ket{\psi_{SR}},\Lambda^{\otimes n}\right),\label{lb}
\end{align}
for any purified quantum state $\ket{\psi_{SR}}$.
Therefore, even though Eq.(\ref{Q(N)}) cannot be solved exactly in many cases, by finding quantum codes with high coherent information, Eq.(\ref{lb}) can be used to get better lower bounds to the quantum capacity of the channel. However, finding such high coherent codes becomes difficult as $n$ increases because the space becomes exponentially large and more complex. Therefore, instead of relying on raw representations and systematic search, one has to switch to some form of variational ansatz and meta-heuristic search mechanism.

In this work, we restrict our study in qubit channels, {\it i.e.} channels whose input and output systems are qubits, $\Lambda:\mathcal{L}(\mathbb{C}^2)\to\mathcal{L}(\mathbb{C}^2)$. We are interested in a $3$-parameter family of such channels called the Pauli channel, and they are defined as,
\begin{align}\label{pauli}
\Lambda(\rho)&=p_0 \sigma_0\rho\sigma_0 + p_1 \sigma_1\rho \sigma_1+p_2 \sigma_2\rho \sigma_2 +p_3 \sigma_3\rho \sigma_3,\\\nonumber
\sigma_0&:=\begin{pmatrix}1&0\\0&1\end{pmatrix},~~~~~~\sigma_1:=\begin{pmatrix}0&1\\1&0\end{pmatrix},\\\nonumber
\sigma_2&:=\begin{pmatrix}0&-\mathtt{i}\\\mathtt{i}&0\end{pmatrix},~~~
\sigma_3:=\begin{pmatrix}1&0\\0&-1\end{pmatrix};\\\nonumber
\mbox{with}&~~p_i\geq0~\forall~i,~~\sum_{i=0}^3p_i=1,~~\&~~\rho\in\mathcal{D}(\mathbb{C}^2).\nonumber
\end{align}
When a quantum state $\rho$ is sent through this channel, physically, the effect can be thought as if the Pauli $\sigma_i$ gate is applied on the system's state with probability $p_i$. We are interested in mapping out the regions in the parameter space where the $2$-shot quantum capacity ($Q^{(1)}(\Lambda^{\otimes 2})$) is provably super-additive. Starting with a neural network ansatz to represent the quantum state $\ket{\psi_{SR}}$ and by following a simple evolutionary strategy for optimization, we found regions in three-parameter space for which coherent information is super-additive. Before dwelling on the details of the result, we quickly review some of the fundamentals of Genetic algorithms.

\subsection{Genetic Algorithms}\label{secIIC}

At its core, machine learning techniques attempt to find patterns hidden in data and to exploit these learned patterns in control and decision making. With the advent of deep learning, complemented by the ever-increasing computational power, machine learning approaches have solved problems that were once thought to be impossible. One of the most striking demonstrations of the power of modern machine learning/AI occurred in 2016 when Google's \textit{AlphaGo} defeated the world champion in the game of \textit{Go}. Apart from this, deep learning techniques have shown exceptional performance in problems like computer vision, natural language processing, medical diagnosis, and more. The success of modern machine learning in such diverse fields has prompted researchers to apply these techniques to fundamental research. Not surprisingly, many physicists have applied machine learning tools in their niche as well \cite{Schuld2015,Biamonte2017,Carleo2019,newcite1,newcite2}. Interestingly, there have also been attempts towards developing methods for machine-assisted discovery of physical principles from experimental data\cite{Iten2020}.
\par
\begin{figure}[t]
\centering
\includegraphics[scale=.3]{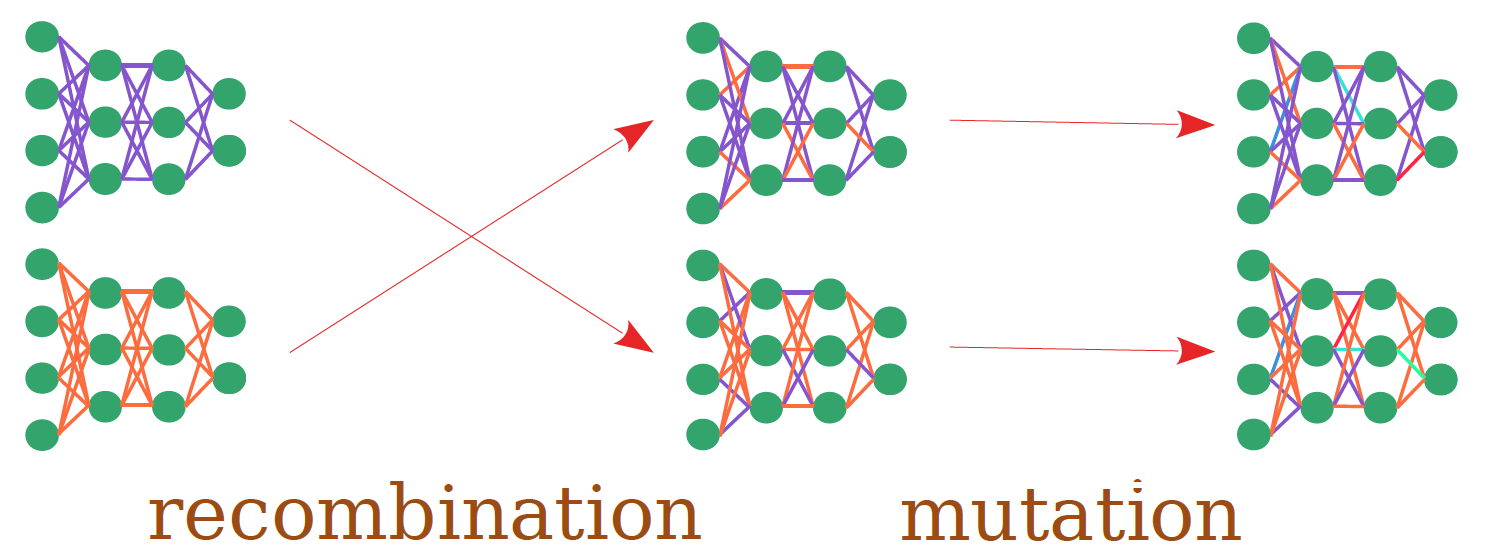}
\caption[Evolution of neural networks]{(Color online) {\bf Evolution of neural networks:} An illustration showing how neural network states evolve to create new generations.}\label{Fig:2}
\end{figure}

 
In this work, we use Neural Networks (NN) to represent quantum states and these NNs are train using Genetic Algorithms. Genetic algorithms (GA) are a class of meta-heuristic search/optimization algorithms inspired by the process of evolution by natural selection. All GAs start with a population of candidate solutions that are evolved over multiple generations to create better (fitter) solutions to the optimization problem. To simulate the various processes in natural selection, there are several biologically inspired operations involved in a GA. The typical GA begins with a randomly initialized set of candidate solutions. The candidate solutions are thought of as individuals of a population. Each individual is assigned a \textit{fitness} that measures the quality of the individual. The population is left to evolve to produce fitter individuals and hopefully to generate a good enough solution to the optimization problem. Typically, the evolution process involves the following three fundamental steps (see Fig. \ref{Fig:2}).
\begin{itemize}
    \item \textbf{Cross-over/recombination}: Several individuals of the population are selected and are crossed over by an appropriately chosen scheme to produce new individuals. Individuals with high fitness are more likely to be crossed over. The intuition is that crossing over can bring together useful features of fit individuals to produce fitter individuals.
    \item \textbf{Mutation}: The new individuals produced through cross-over are subjected to mutation. Mutation introduces more variations into the population.
    \item \textbf{Selection}: The fitter individuals are promoted to the next generation while individuals with less fitness are removed from the population with high probability.
\end{itemize}
GAs are derivative-free optimization methods, making these particularly useful for optimizing discontinuous objective functions and landscapes filled with local minima. Since GAs are good at avoiding local optima in complex optimization landscapes, GAs are often the tool of choice for global optimization problems. All this said, GAs have their limitations. Since information like the gradient is not considered in the optimization process, GAs can be resource hungry, requiring large population sizes and many fitness function evaluations. This can result in a substantial computational overhead if the fitness function evaluation is computationally expensive. In such cases, one often has to resort to parallel computing or computationally efficient approximations of the fitness function.

\subsection{Methods}\label{secIID}
\begin{figure}[t!]
\centering
\includegraphics[scale=0.5]{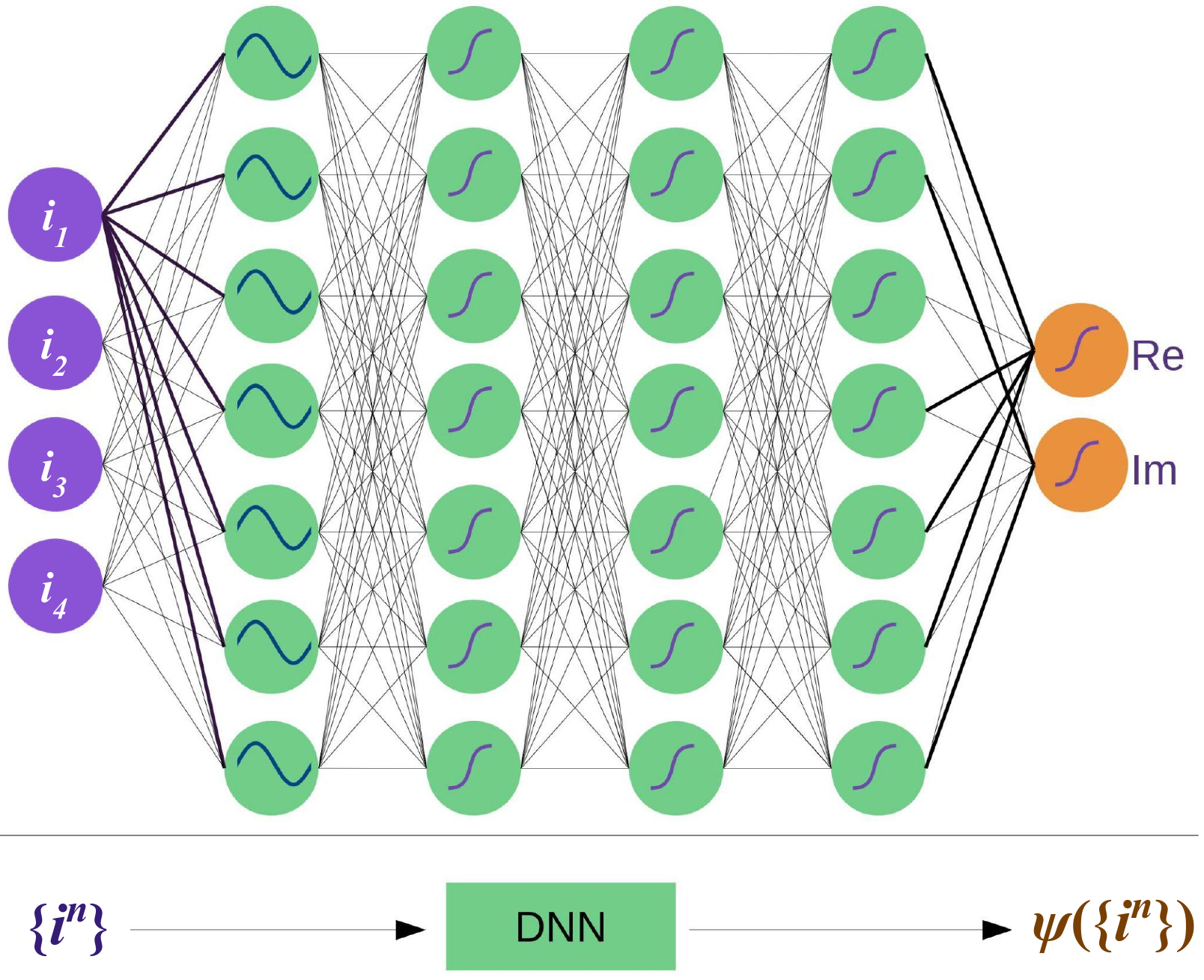}
\caption[The Neural network ansatz]{(Color online) {\bf The Neural Network ansatz:}A simple illustration of neural network ansatz used in this study. The given NN represents a 4-qubit quantum state. The inputs given to the NN are binary strings of length four and represent the corresponding computational basis states. The two outputs from the NN represent the real and imaginary parts of the amplitude corresponding to the input basis state. The first layer has a cosine activation, while all subsequent layers are given a tanh activation function.}\label{Fig:3}
\end{figure}

We represented quantum states using a neural network (NN) ansatz \cite{NNState1,Saito2017} as shown in Fig. \ref{Fig:3}. More specifically, this is a Feed Forward (FF) neural network, which we chose following it's superior performance in bench-marking problems compared to other NN architectures like Restricted Boltzman Machines, as reported in Ref.\cite{Bausch2020}. If there are n-qubits in the problem, a pure quantum state $\ket{\Psi}$ can be expanded in the computational basis as,
\begin{align}\label{ansatz}
\ket{\Psi}=\sum_{\{i^n\}}\frac{1}{C}\psi(\{i^n\})\ket{\{i^n\}},
\end{align}
$\psi(\{i^n\})$ is the complex amplitude of the computational basis state $\ket{\{i^n\}}$ identified by the bit string $\{i^n\}$. C is the normalization factor. In the NN ansatz, the state $\ket{\Psi}$ is represented using a NN with n-input neurons and two output neurons. When fed in with a bit string $\{i^n\}$, the NN outputs two values corresponding to the real and imaginary parts of $\psi(\{i^n\})$. The weights of the NNs are the variational parameters in this representation that may be varied using an appropriate optimization technique. There are no formal rules about the choice of activation functions; however, a cosine activation for the first hidden layer and tanh activation for all subsequent hidden layers are known to work very well in practice \cite{Bausch2020,NNState1}, and it is employed in this work. We implemented the fully connected deep neural network architecture using the Sequential API of the TensorFlow package \cite{Abadi2015}.

An immediate question that follows is whether using such a NN ansatz is needed in the first place. Why not use a raw encoding (ansatz), where the individual represents the coefficients of the statevector in a chosen basis, directly without going through the pain of introducing more parameters in the form of a neural network? This question has already been answered in {Ref.\cite{Bausch2020}}, where they provided empirical evidence that showed the superior performance of the NN ansatz compared to raw encodings. For completeness, we have also analysed the performance of the NN ansatz and found that for the two-shot problem, using a NN ansatz might be an overkill, but in the three-shot case, the NN ansatz significantly outperforms the raw encoding. See the Appendix \ref{Appendix} for more details.

To fully define the optimization problem, we need a loss function (the function to be optimized). We formulate the problem as a minimization problem, and hence the loss function is simply the negative of the coherent information. Initially, we tried several gradient-based optimizers implemented in TensorFlow, including the standard 'Adam' optimizer \cite{Kingma2014}. These techniques worked well in 2-qubit space, but as the space became large, all gradient-based algorithms almost always converged to some local minima with vanishing coherent information. This is expected since in the noisy regime, regions of positive, coherent information are likely to cover only a small fraction of the Hilbert space volume, and all product states ($\ket{\Psi_{SR}}=\ket{\psi_{S}}\otimes\ket{\phi_{R}}$) form local minima with vanishing coherent information. One trick to get around this problem is to add bias into the search space by penalizing product states. However, our analysis suggested that this is not a very robust approach. 
\par
When the optimization landscape is filled with local optima, a better strategy is to use some kind of gradient free-optimization methods like Particle Swarm optimization (PSO) or GAs. Ref.\cite{Bausch2020}
studied both PSO and a simple GA in benchmark problems and found that both methods shows comparable performance.
 
In this work, we choose to use a simple Genetic Algorithm as presented in Ref. \cite{back2018evolutionary}. The algorithm comes prebuilt with the DEAP (Distributed Evolutionary Algorithms in Python) package \cite{Fortin2012} as {\textit{deap.algorithms.eaSimple}}. In addition to being an easy to use package, DEAP implementations are also easily parallelizable. For completion, we have also made detailed comparison of the performance of the GA scheme we employed to a simple PSO variant, concluding that for the problem at hand both schemes are comparable in terms of performance. We have also observed some interesting nuances in the behaviour of these algorithms with variations in depth of the NN and other meta-parameters. Detailed discussions can be found in the Appendix \ref{Appendix}.
\par
Individuals (candidate solutions) in the population are represented as a list consisting of weight matrices of the Neural Network. We employed a recombination scheme such that each of these weight matrices independently undergo a $2D$ cross-over mechanism {\cite{Cohoon1987}}, where randomly chosen sub-matrices of each weight matrix is swapped with the corresponding sub-matrix of the other parent individual. In some sense, these weight matrices are similar to chromosomes in that recombination (cross-over) happens only between weights of the corresponding layers of two individuals. 
 
We also added a Gaussian mutation of predetermined mean and standard deviation to the new individuals produced after cross-over. Fig. \ref{Fig:2} gives a simplified illustration of the cross-over and mutation process described above. We now have all the machinery necessary to find quantum codes with high coherent information for the Pauli channel.

\section{Results and Discussion}\label{secIV}

\subsection{One-shot capacity of the Pauli Channel}
To prove the super-additivity of coherent information for a channel, it is necessary that one knows its one-shot quantum capacity ($Q^{(1)}(\Lambda)$) exactly. For the depolarising channel, which is a special case of the class of Pauli channels, it is known that the maximally entangled state maximizes the single-shot coherent information \cite{wilde}. It turns out that this result also carries over to the class of Pauli channels as shown in Ref. \cite{pauli_single_shot,Bennett96}. Thus for a Pauli channel $\Lambda_{\vec{p}}$ as in Eq.(\ref{pauli}), the one-shot quantum capacity is given by,
\begin{align}
    Q^{(1)}(\Lambda_{\vec{p}} ) &=1-H(\vec{p})
\end{align}
where, $ \vec{p}=(p_0,p_1,p_2,p_3)$ is the probability vector and $H(.)$ is the Shannon entropy of the probability distribution. 

The {\it quantum code} that maximises the single-shot coherent information is the maximally entangled state given by,

\begin{align}
\ket{\Psi_{SR}}=\frac{1}{\sqrt{2}}(\ket{0}_R\otimes\ket{0}_S+\ket{1}_R\otimes\ket{1}_S).
\end{align}
Fig. \ref{Fig:5} shows coherent information of the maximally entangled state for Pauli channels with parameters in the range, $0<p_i<0.2,~i\in\{1,2,3\}$. As seen from the figure, coherent information is high and positive in the low-noise regime. As the channel gets noisier, the single-shot coherent information quickly reduces to zero. We have explored this three-parameter space for the two channel-use and three channel-use case to find quantum codes that display super-additivity of coherent information.

\subsection{Results of the GA based optimization}
First we will consider the problem of finding good quantum codes with high two-shot coherent information. After analysing results, presenting regions with super-additivity and the quantum codes with high two-shot coherent information, we move to the case of three-shot coherent information.
\subsubsection{Optimising two-shot coherent information}
Quantum states were represented using a fully connected neural network with a 4-neuron input layer, 4 hidden layers with a width of 4 neurons each, and a 2-neuron output layer. The first hidden layer has a cosine activation function, and others have a `tanh' activation function. All optimizations were done using a simple evolutionary algorithm (described in \cite{back2018evolutionary}) as implemented in the DEAP package\cite{DEAP}. We set the cross-over probability to $0.5$, mutation probability to $0.2$, and a population consisting of a total of $300$ individuals evolved for $300$ generations. Mutations were modeled as a simple Gaussian addition with a mean of $0.5$ and a standard deviation of $0.25$. Each attribute was allowed to mutate with a probability of $0.5$. The selection operation was based on a tournament selection scheme with three individuals participating in each tournament. The averaged out learning curves of the algorithm with different meta-parameter settings can be seen in the Appendix \ref{Appendix}.
\par
Optimization results were very promising with $\frac{1}{2}\mathcal{I}(\ket{\Psi},\Lambda^{\otimes 2})$ almost always reaching at least as high as the one-shot quantum capacity, where $\ket{\Psi}\in(\mathbb{C}^4)^{\otimes2}\equiv\mathbb{C}^2_{S_1}\otimes\mathbb{C}^2_{S_2}\otimes\mathbb{C}^2_{R_1}\otimes\mathbb{C}^2_{R_2}$. This is a good indication that the optimization algorithm is covering our desire region of the $4$-qubit space and not getting stuck in most local optima. Interestingly, in many points in the search space, the algorithm found quantum codes $\ket{\Psi}\in(\mathbb{C}^4)^{\otimes2}$ with coherent information strictly greater than one-shot capacity, i.e.,
\begin{align}
\frac{1}{2}\mathcal{I}(\ket{\Psi},\Lambda^{\otimes 2}(p_1,p_2,p_3))>Q^{(1)}(\Lambda(p_1,p_2,p_3));\label{ineq}
\end{align}
and hence establishing super-additivity of coherent information for those channels.

 \begin{figure}[t]
    \subfloat[Population size:50]{%
      \includegraphics[width=0.45\textwidth]{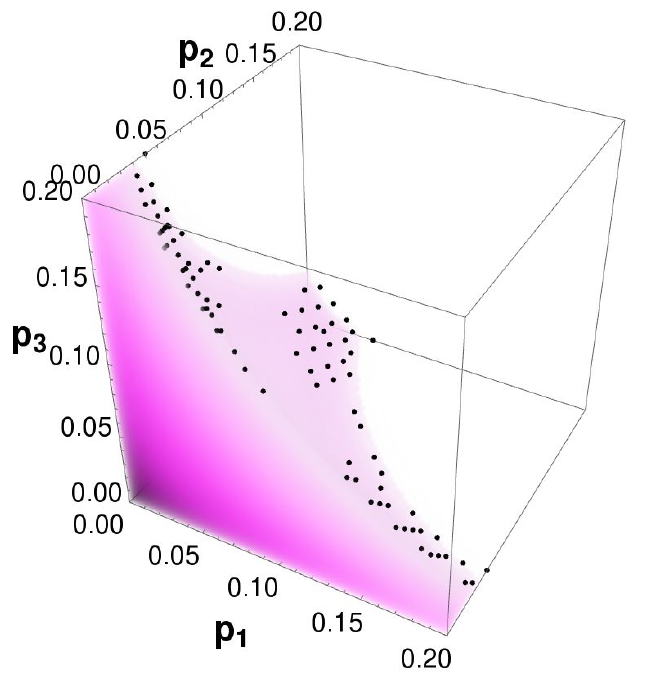}
    }
    \\
    \subfloat[Population size:300]{%
      \includegraphics[width=0.45\textwidth]{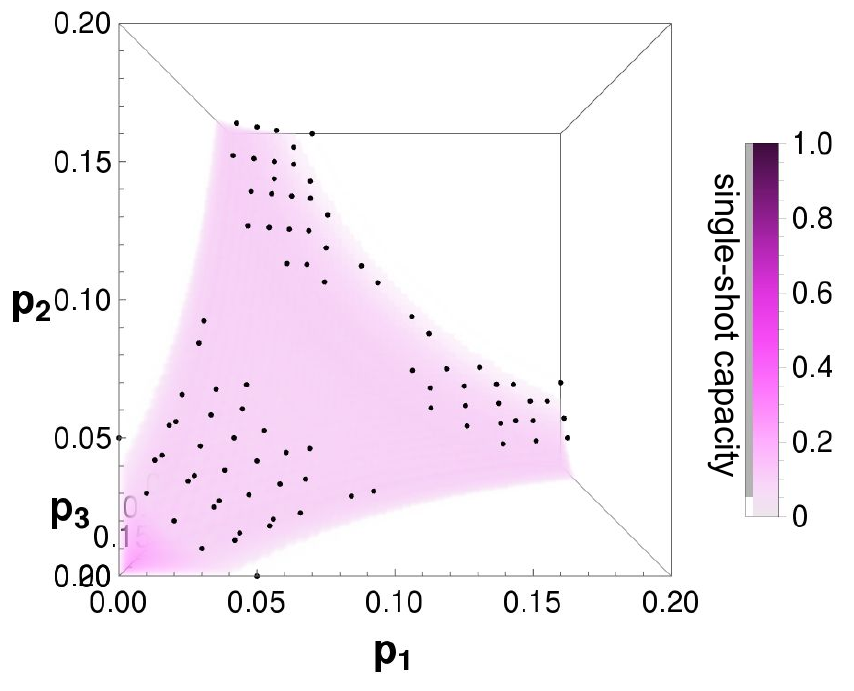}
    }
    \caption[Result of GA optimsation]{(Color online) {\bf Result of GA optimization :} The single-shot capacity of the class of Pauli channels are shown in the density plot, color-coded as detailed in the colorbar. Each point in the $p_1p_2p_3$ space represents a Pauli channel. We studied the super-additivity of points in a 3D grid of width 0.01. The black spots in the graph represents Pauli channels that display super-additivity of coherent information among these points.}\label{Fig:5}
  \end{figure}

The set of Pauli channels for which the algorithm found super-additive quantum codes are shown as black spots in Fig. \ref{Fig:5}, along with the channel's one-shot capacity. As can be seen from the figure, there appears to be an apparent symmetry in the regions with super-additivity, with respect to the permutations of the parameters $p_1,~p_2,~\&~p_3$. As expected, we could not find any super-additivity for two channel-use along the main diagonal; the diagonal represents the family of depolarizing channels. To cross-check this observation, we ran a more refined optimization for the depolarizing channel for the two channel-use case, with a population of $500$ individuals that evolved for $1000$ generations; even then, the algorithm could not find any super-additive quantum codes. Therefore, this observation suggests that, most likely, two channel-uses cannot result in super-additivity for the family of depolarizing channels.
\begin{figure}[t!]
\centering
\includegraphics[scale=0.5]{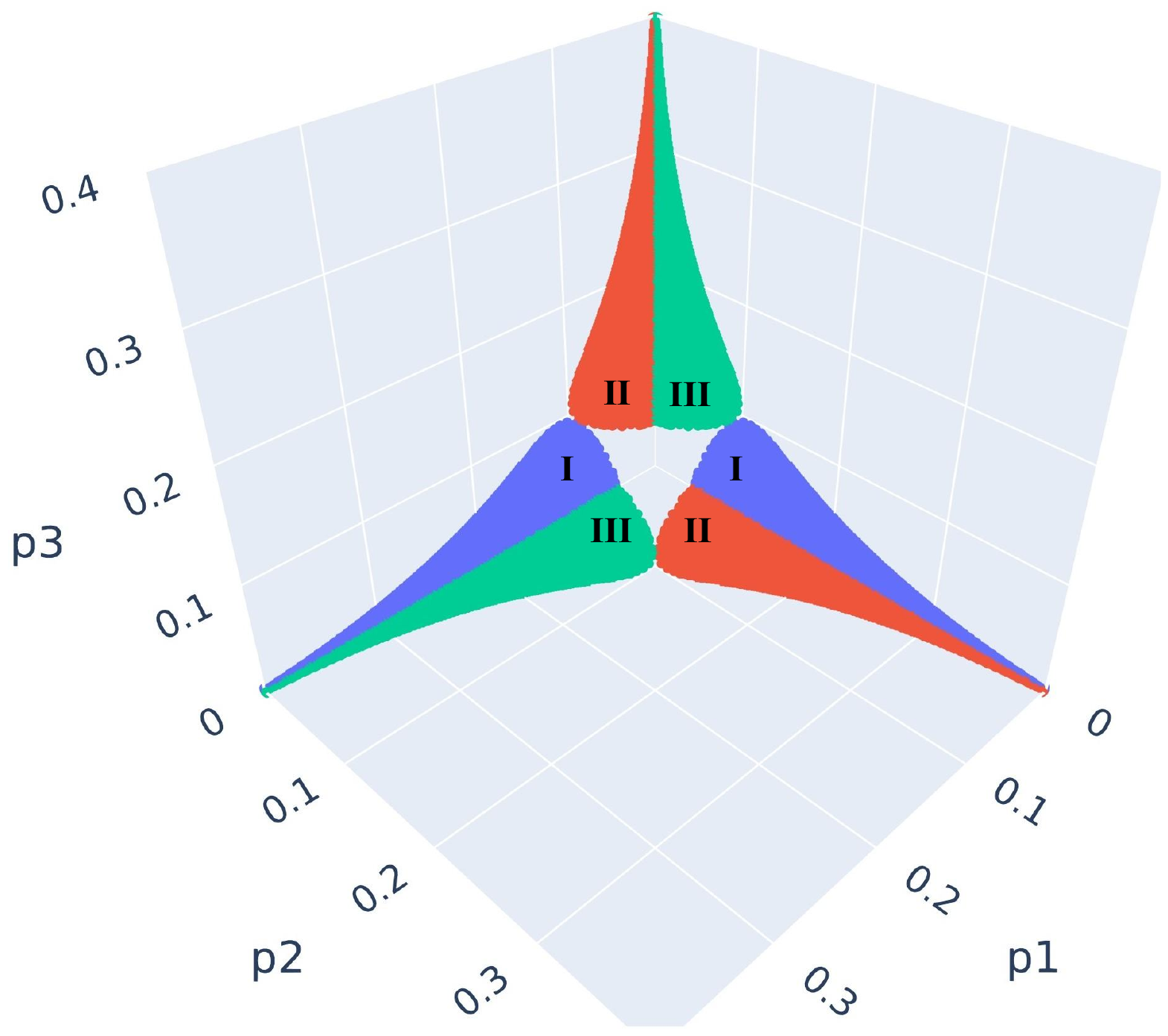}
\caption[Result of GA optimization]{(Color online) {\bf Regions of two-shot super-additivity obtained from the three quantum codes :} Blue region (marked with `I') shows highest two-shot coherent information when code $\ket{\Psi(I)}$ is used. Similarly, Red region (marked `II') corresponds to code $\ket{\Psi(II)}$ and Green region (marked `III') corresponds to code $\ket{\Psi(III)}$. $p_1,p_2,p_3$ are respectively the probabilities of three Pauli errors. }\label{Fig:6}
\end{figure}

The next immediate step is to find the quantum codes that showed super-additivity in various regions; for all we know, the quantum states learned may vary from point-to-point displaying a non-trivial dependence on the three parameters. But, surprisingly, after looking at the numerical value of the states and accounting for equivalences upto local unitaries in the puryifing qubits, we found that the algorithm found only three distinct quantum codes that display super-additivity of coherent information. Moreover, the regions of super-additivity (shown in Fig. \ref{Fig:6}) separate into three distinct regions, depending on which quantum code gives maximal separation between one-shot and two-shot quantum capacity. The three (unnormalized) quantum codes the algorithm learned are given by,
\begin{equation}
\begin{aligned}
\ket{\Psi(I)}&\equiv\ket{zz}\ket{\bar{z}z}+\ket{zz}\ket{\bar{z}\bar{z}}+\ket{\bar{z}\bar{z}}\ket{zz}+\ket{\bar{z}\bar{z}}\ket{z\bar{z}};\\
\ket{\Psi(II)}&\equiv\ket{xx}\ket{\bar{x}x}+\ket{xx}\ket{\bar{x}\bar{x}}+\ket{\bar{x}\bar{x}}\ket{xx}+\ket{\bar{x}\bar{x}}\ket{x\bar{x}};\\
\ket{\Psi(III)}&\equiv\ket{yy}\ket{\bar{y}y}+\ket{yy}\ket{\bar{y}\bar{y}}+\ket{\bar{y}\bar{y}}\ket{yy}+\ket{\bar{y}\bar{y}}\ket{y\bar{y}};\\
\end{aligned}\label{Quantumcodes3}
\end{equation}
where, $\ket{a}~\left(\ket{\bar{a}}\right)$ is the $+1~\left(-1\right)$ eigenstate of the Pauli $\sigma_a$ operator, $a\in\{x,y,z\}$. Note that $\ket{\Psi(I)}$ is invariant under $ZZ$ operator on the first two qubits. Similarly, the codes $\ket{\Psi(II)}$ and $\ket{\Psi(III)}$ are invariant under the action of $XX$ and $YY$, respectively. The expression for two-shot quantum capacity when $\ket{\Psi(I)}$ is passed through the Pauli channel $\Lambda_{\vec{p}}\equiv\Lambda(p_1,p_2,p_3)$, with $\vec{p}=(p_1,p_2,p_3)$ and $p_0=1-p_1-p_2-p_3$ is given by,
\begin{widetext}
\begin{equation}
\begin{aligned}
    Q^{(2)}(\ket{\Psi(I)},\Lambda_{\vec{p}})=&p_1 p_2 \log \left(2 p_1 p_2\right)+p_0 p_3 \log \left(2 p_0 p_3\right)+\frac{1}{2}\left(p_1^2+p_2^2\right) \log \left(p_1^2+p_2^2\right)+\frac{1}{2}\left(p_0^2+p_3^2\right)
   \log \left(p_0^2+p_3^2\right)\\
    &-\left(p_1+p_2\right) \left(p_0+p_3\right) \log 
   \left(\left(p_1+p_2\right) \left(p_0+p_3\right)\right)
   +\left(p_0 p_2+p_1 p_3\right) \log \left(p_0 p_2+p_1 p_3\right)\\
   &+\left(p_0 p_1+p_2 p_3\right) \log \left(p_0 p_1+p_2 p_3\right)-\frac{1}{2}\left(p_0+p_3\right){}^2 \log \left(\frac{1}{2} \left(\left(p_1+p_2\right){}^2+\left(p_0+p_3\right){}^2\right)\right)\\
   &-\frac{1}{2}\left(p_1+p_2\right){}^2 \log \left(\frac{1}{2} \left(\left(p_1+p_2\right){}^2+\left(p_0+p_3\right){}^2\right)\right)\label{q22}
\end{aligned}
\end{equation}
\end{widetext}
\normalsize
Given Eq.(\ref{q22}), symmetry of the quantum codes and that of the Pauli channel can be used to show that
\footnotesize
\begin{align*}
    Q^{(2)}(\ket{\Psi(II)},\Lambda(p_1,p_2,p_3))&=Q^{(2)}(\ket{\Psi(I)},\Lambda(p_3,p_2,p_1))\\
    Q^{(2)}(\ket{\Psi(III)},\Lambda(p_1,p_2,p_3))&=Q^{(2)}(\ket{\Psi(I)},\Lambda(p_1,p_3,p_2))
\end{align*}
\normalsize
The regions in the three-parameter space of $p_1,p_2$ and $p_3$ that shows super-additivity using these codes are depicted in Fig. \ref{Fig:6}. It should be pointed out that there are overlaps between regions that shows super-additivity, for instance $\ket{\Psi(I)}$ also satisfies Eq.(\ref{ineq}) in certain regions in Region:II and Region:III closer to the $p_2p_1$-plane, but these values are less than what one can achieve from $\ket{\Psi(II)}$ and $\ket{\Psi(III)}$ respectively. 

\begin{figure}[t!]
\centering
\includegraphics[scale=0.60]{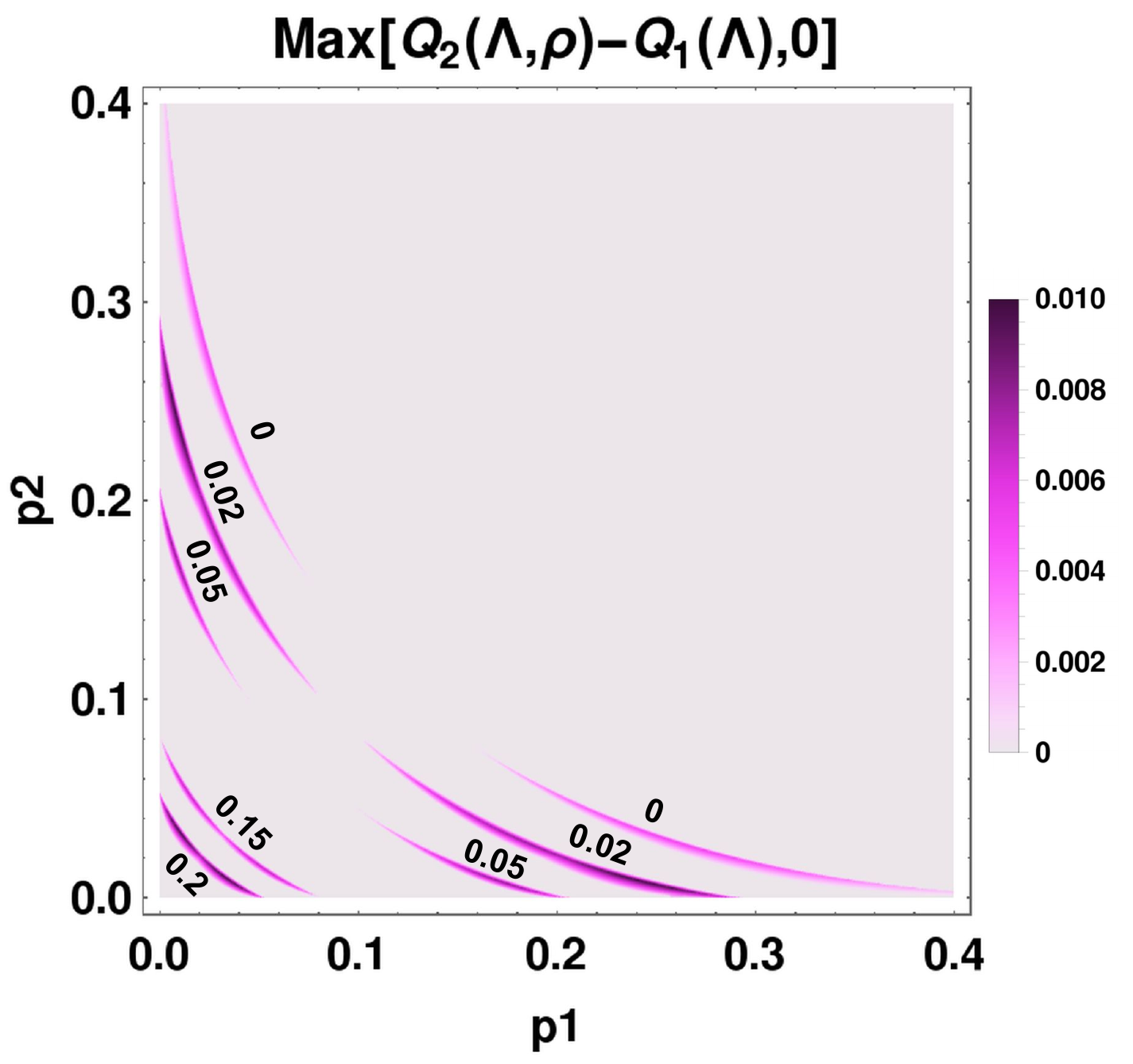}
\caption[Result of GA optimization]{(Color online) {\bf 2D Density plot of Super-additivity for fixed $p_3$: }Max[$Q^{(2)}(\Psi,\Lambda_{\vec{p}})-Q^{(1)}(\Lambda_{\vec{p}})$,0] is plotted for fixed values of $p_3\in\{0,0.02,0.05,0.15,0.2\}$, where $Q^{(2)}(\Psi,\Lambda_{\vec{p}})\equiv\max[Q^{(2)}(\ket{\Psi(I)},\Lambda_{\vec{p}}),Q^{(2)}(\ket{\Psi(II)},\Lambda_{\vec{p}}),Q^{(2)}(\ket{\Psi(III)},\Lambda_{\vec{p}})]$. All coloured regions corresponds to regions of superadditivity, with the amount of superadditivity shows by the `colorbar' to the right of the plot. }\label{Fig:7}
\end{figure}

Fig. \ref{Fig:6} gives the general structure of regions where super-additivity has been shown, but it tells very little about the three-dimensional structure of the regions that display super-additivity and the magnitude of super-additivity. To get around this problem, two-dimensional density plots were created, as shown in Fig. \ref{Fig:7}, for different fixed values of $p_3$ ($\in\{0,0.02,0.05,0.15,0.2\}$). The colored regions in the plot represent regions of super-additivity, with the magnitude of super-additivity encoded in the color as detailed in the colorbar. Superadditivity of two-shot coherent information has been shown for several two-parameter families of quantum channels like Dephrasure channel and the Generalised Amplitude Damping Channel\cite{Bausch2020}. Each stripe in Fig. \ref{Fig:7} effectively describes another two-parameter family with parameters $p_1$ and $p_2$ which shows superadditivity of two-shot coherent information. Fig. \ref{Fig:6} coupled with Fig. \ref{Fig:7} gives a better description of the regions where super-additivity can be shown and the magnitude of this super-additivity. From an experimentalist's perspective, it might be useful to find Pauli channels that exhibit maximal gap between the one-shot quantum capacity and two-shot quantum capacity while using the codes Eq.(\ref{Quantumcodes3}). Numerically, we calculated the maximal gap to be $\max_{\vec{p}}[Q^{(2)}(\Lambda_{\vec{p}},\ket{\Psi(I)})-Q^{(1)}(\Lambda_{\vec{p}})]\approx 0.0102342$, achieved using the Pauli channel with $\vec{p}=(p_1,p_2,p_3)=(0.225688, 0.00801196, 0.0263041)$ (and other permutations of $p_1$, $p_2$ and $p_3$). This is, to the best of our knowledge, much higher than the gap obtained in other well known channels such as the generalized erasure channel\cite{Filippov_2021}, Dephrasure channel, Generalised Amplitude Damping Channel, and the Depolarising channel\cite{Bausch2020}, making such Pauli channels good candidates for experimental studies.

\begin{figure}[t!]
\centering
\includegraphics[scale=0.55]{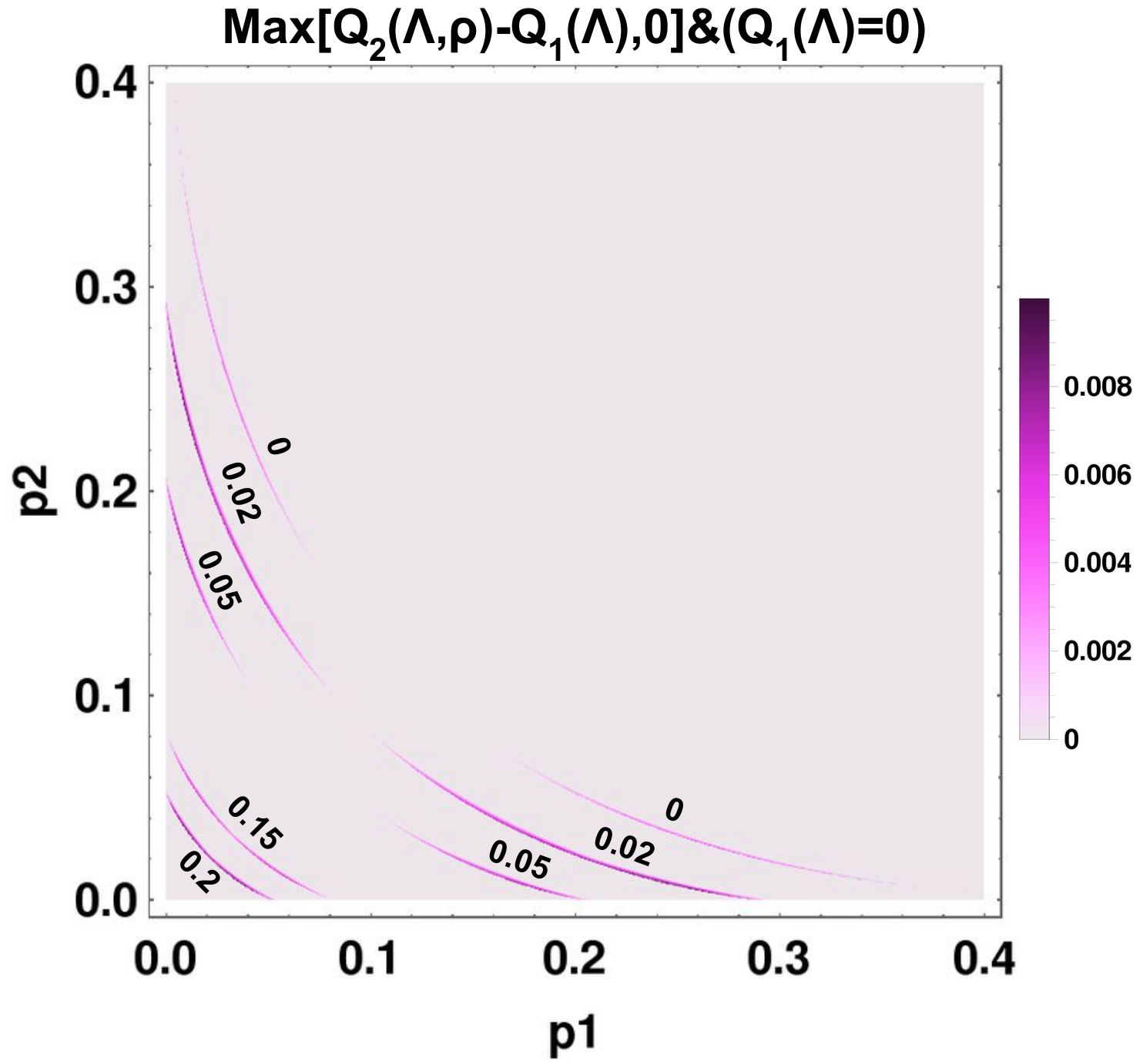}
\caption{(Color online) {\bf Pauli channels with $Q^{(1)}(\Lambda_{\vec{p}})=0$ but $Q^{(2)}(\Lambda_{\vec{p}},\rho) > 0 $:} Max[$Q^{(2)}(\lambda_{\vec{p}},\rho)-Q^{(1)}(\Lambda_{\vec{p}})$,0] is plotted for fixed values of $p_3\in\{0,0.02,0.05,0.15,0.2\}$, whenever $Q^{(1)}(\Lambda_{\vec{p}})=0$. All coloured regions corresponds to regions of superadditivity, with the amount of superadditivity shows by the `colorbar' to the right of the plot. }\label{Fig:8}
\end{figure}

It is also interesting to find out sub-classes of Pauli channels for which $Q^{(1)}(\Lambda_{\vec{p}})=0$ but $Q^{(2)}(\Psi,\Lambda_{\vec{p}}) \ne 0$. These are the channels where, on single-use of the channel, no quantum information can be communicated, but on multiple parallel uses, communication is viable. Pauli channels that show this property are shown in Fig. \ref{Fig:8}. This region is identical to that of Fig. \ref{Fig:7}, except that it is `thinner' than that in Fig. \ref{Fig:7}.

\subsubsection{Optimising three-shot coherent information}
\begin{figure}[t!]
\centering
\includegraphics[scale=0.5]{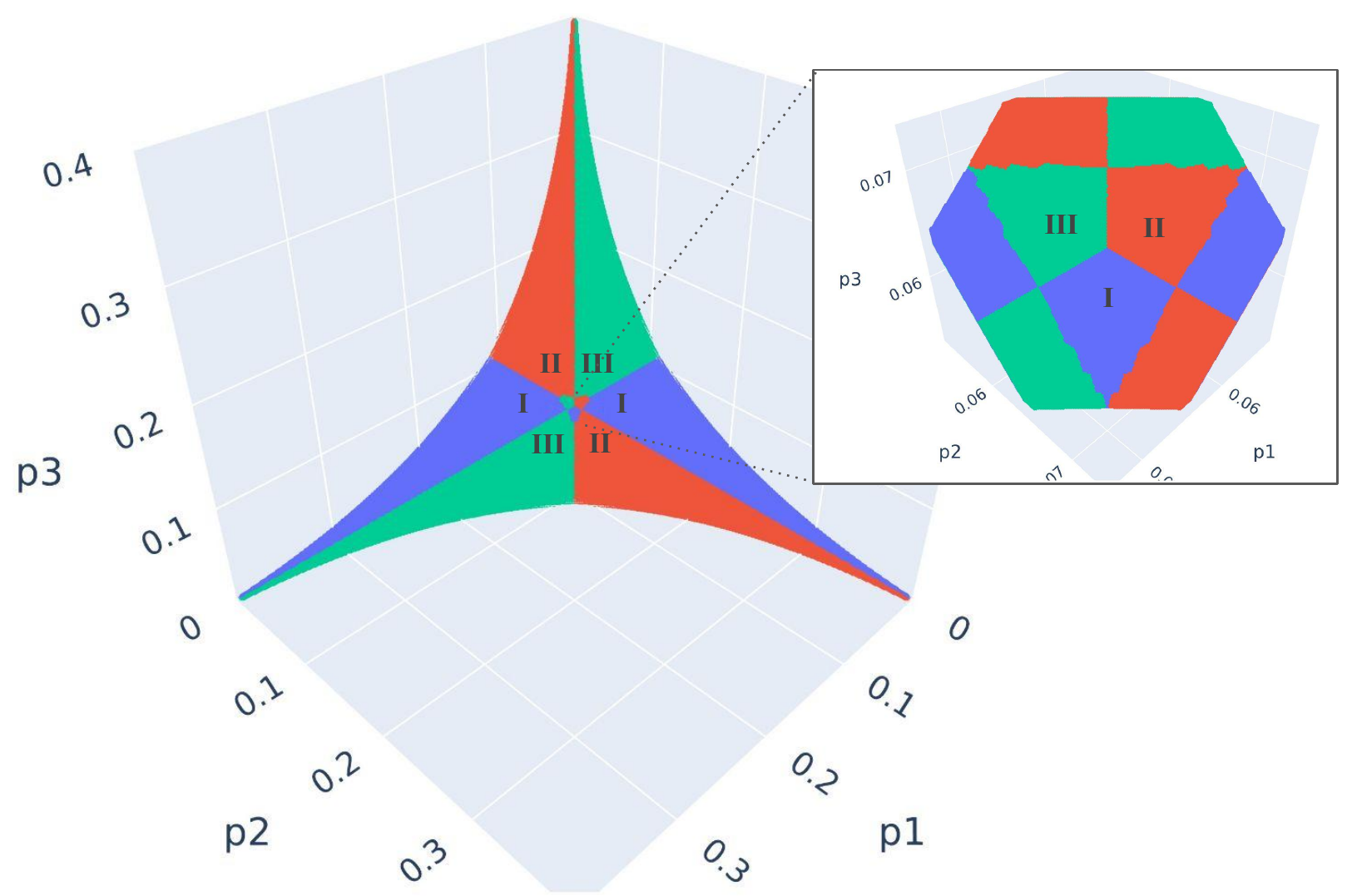}
\caption{(Color online) {\bf Regions of three-shot super-additivity obtained from the three quantum codes:} The Blue region (marked with `I') shows the highest three-shot coherent information when code $\ket{\Phi(I)}$ is used. Similarly, Red region (marked `II') corresponds to code $\ket{\Phi(II)}$ and Green region (marked `III') corresponds to code $\ket{\Phi(III)}$. $p_1,p_2,p_3$ are respectively the probabilities of three Pauli errors. }\label{Fig:9}
\end{figure}

Till now, we restricted our discussions to the two-shot capacity. It can be seen that the depolarizing channel (corresponding to the main diagonal in Fig. \ref{Fig:7}) does not show superadditivity under two channel uses using the quantum codes shown in Eq.(\ref{Quantumcodes3}). In fact, no quantum codes are known that achieve superadditivity for the two-shot case for the depolarising channel. However, it is known that by using the repetition code, the depolarising channel shows superadditivity under three channel-use (and higher uses as well) \cite{Bausch2020}. It is, therefore, interesting to study the three-shot capacity of the Pauli channels. Using GA-based search with a Schmidt  ansatz\cite{Bausch2020} for representing quantum states, we searched through the three-parameter space of Pauli channels for the super-additivity phenomenon, finding the regions marked in Fig.\ref{Fig:9}. We found three distinct quantum codes that displayed the highest coherent information in our analysis. These are,
\begin{equation}
\begin{aligned}
\ket{\Phi(I)}&\equiv\ket{zzz}\ket{z}+\ket{\bar{z}\bar{z}\bar{z}}\ket{\bar{z}};\\
\ket{\Phi(II)}&\equiv\ket{xxx}\ket{x}+\ket{\bar{x}\bar{x}\bar{x}}\ket{\bar{x}};\\
\ket{\Phi(III)}&\equiv\ket{yyy}\ket{y}+\ket{\bar{y}\bar{y}\bar{y}}\ket{\bar{y}} 
\end{aligned}\label{three}
\end{equation}
where the ket symbols are defined as earlier. These are essentially the famous repetition codes which have been used to show super-additivity for the depolarizing channels. The last qubit in equation set.\ref{three} is the purifying qubit. The regions in the three-parameter space that shows super-additivity using these codes are shown in Fig. \ref{Fig:9}. The regions look similar to that in Fig. \ref{Fig:6} for the two-shot case, but we emphasize that although there is an overlap between the two regions in Fig. \ref{Fig:9} and Fig. \ref{Fig:6}, neither of the regions fully contain the other. It is clear from Fig. \ref{Fig:9} that there are Pauli channels of the form $\vec{p}=(p_1,p_2,p_3)=((1+\epsilon_1) p,(1+\epsilon_2) p,(1+\epsilon_3) p)$ with sufficiently small $\epsilon_i$ that does not show superadditivity under two channel uses(using the quantum codes we described), but does show superadditivity under three uses. The depolarising channel is, of course, a special case of these channels. It is also interesting to note that there is a flip in the good quantum codes near the centre. This again reveals the non-trivial relation between noise and `goodness' of quantum codes. The exact reason for this behaviour is an interesting subject for future study.

We further calculated the expression for three-shot quantum capacity when $\ket{\Phi(I)}$ is passed through the Pauli channel $\Lambda_{\vec{p}}\equiv\Lambda(p_1,p_2,p_3)$, with $\vec{p}=(p_1,p_2,p_3)$ and $p_0=1-p_1-p_2-p_3$. It can be written as,
\begin{widetext}
\begin{equation}
    \begin{aligned}
Q^{(3)}(\ket{\Phi(I)},\Lambda_{\vec{p}})&=\frac{1}{3} p_2 \left(3 p_1^2+p_2^2\right) \log \left(p_2^3+3 p_1^2
   p_2\right)+\frac{1}{3} p_1 \left(p_1^2+3 p_2^2\right) \log \left(p_1^3+3 p_2^2
   p_1\right)\\
   &-\left(p_1+p_2\right) \left(p_0+p_3\right) \log \left(\frac{1}{2}
   \left(p_1+p_2\right) \left(p_0+p_3\right)\right)\\
   &~~~+\left(p_0
   \left(p_1^2+p_2^2\right)+2 p_1 p_2 p_3\right) \log \left(p_0
   \left(p_1^2+p_2^2\right)+2 p_1 p_2 p_3\right)\\
   &+\left(2 p_0 p_1
   p_2+\left(p_1^2+p_2^2\right) p_3\right) \log \left(2 p_0 p_1
   p_2+\left(p_1^2+p_2^2\right) p_3\right)\\
   &+\frac{1}{3} p_3 \left(3
   p_0^2+p_3^2\right) \log \left(p_3^3+3 p_0^2 p_3\right)+\frac{1}{3} p_0
   \left(p_0^2+3 p_3^2\right) \log \left(p_0^3+3 p_3^2 p_0\right)\\
   &-\frac{1}{3}
   \left(1-3 \left(p_1+p_2\right) \left(p_0+p_3\right)\right) \log
   \left(\frac{1}{2} \left(1-3 \left(p_1+p_2\right)
   \left(p_0+p_3\right)\right)\right)\\
   &+\left(2 p_0 p_2 p_3+p_1
   \left(p_0^2+p_3^2\right)\right) \log \left(2 p_0 p_2 p_3+p_1
   \left(p_0^2+p_3^2\right)\right)\\
   &+\left(2 p_0 p_1 p_3+p_2
   \left(p_0^2+p_3^2\right)\right) \log \left(2 p_0 p_1 p_3+p_2
   \left(p_0^2+p_3^2\right)\right).\label{q33}
 \end{aligned}
\end{equation}
\end{widetext}
\normalsize

Like in the previous case, given Eq.(\ref{q33}), symmetry of the quantum codes and that of the Pauli channels can be used to show that,
\small
\begin{align*}
    Q^{(3)}(\ket{\Phi(II)},\Lambda(p_1,p_2,p_3))&=Q^{(3)}(\ket{\Phi(I)},\Lambda(p_3,p_2,p_1))\\
    Q^{(3)}(\ket{\Phi(III)},\Lambda(p_1,p_2,p_3))&=Q^{(3)}(\ket{\Phi(I)},\Lambda(p_1,p_3,p_2))
\end{align*}
\normalsize

Here as well, we numerical estimated the maximal gap while using the quantum codes in set of Eq.(\ref{q33}) to be $\max_{\vec{p}}[Q^{(3)}(\Lambda_{\vec{p}},\ket{\Phi(I)})-Q^{(1)}(\Lambda_{\vec{p}})]\approx 0.0127406$, achieved using the Pauli channel with $\vec{p}=(p_1,p_2,p_3)=(0.00730649,0.240303, 0.0223234)$ (and other permutations of $p_1$, $p_2$ and $p_3$). Interestingly, the optimal gap in this case is much higher than what was observed in the case of two channel uses discussed earlier. 

\begin{figure}[t!]
\centering
\includegraphics[scale=0.6]{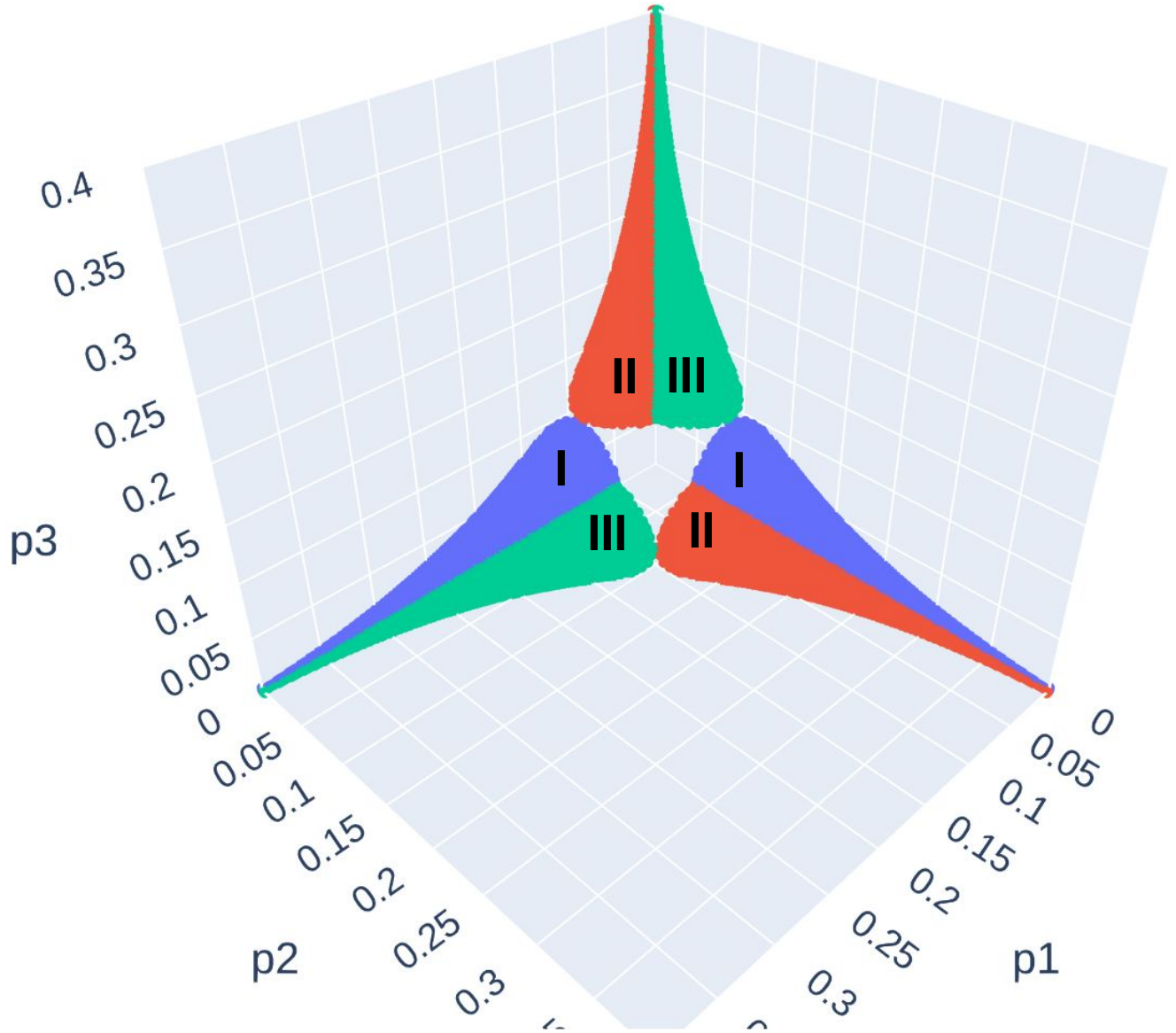}
\caption[Result of GA optimization]{(Color online) {\bf Regions of three-shot super-additivity obtained from the quantum codes in Eq.(\ref{four}):} Blue region (marked with `I') shows highest three-shot coherent information when code $\ket{\chi(I)}$ is used. Similarly, Red region (marked `II') corresponds to code $\ket{\chi(II)}$ and Green region (marked `III') corresponds to code $\ket{\chi(III)}$. $p_1,p_2,p_3$ are respectively the probabilities of three Pauli errors. }\label{Fig:10}
\end{figure}

\begin{figure}[t!]
\centering
\includegraphics[scale=0.6]{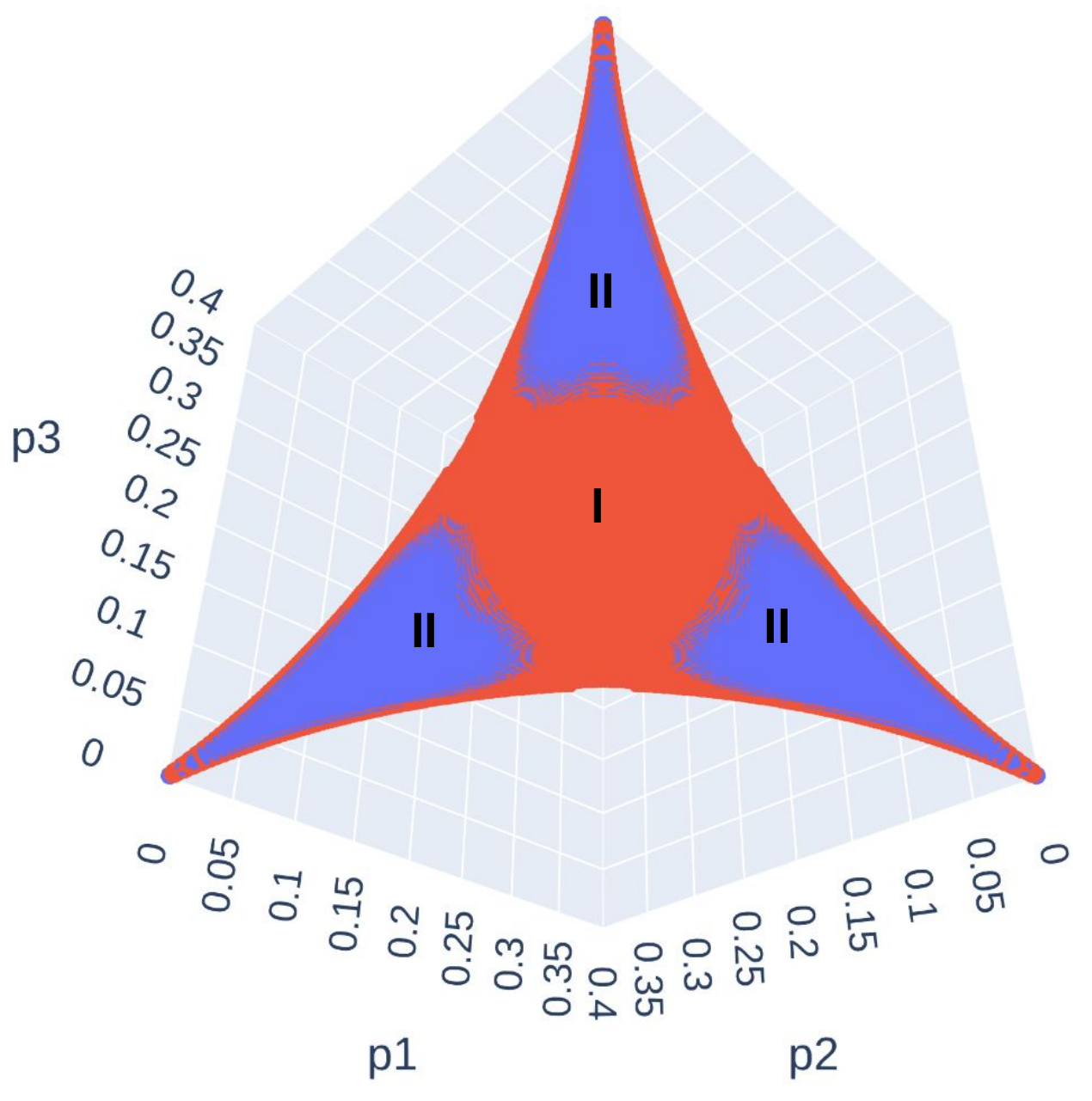}
\caption[Region of 3-shot Super-additivity]{(Color online) {\bf Regions of three-shot super-additivity obtained from the quantum codes in Eq.({\ref{four}}) and Eq.(\ref{three}}): The red region (marked 'I') corresponds to points where Eq.(\ref{three}) shows maximum Super-additivity while the blue region (marked 'II') corresponds to points where Eq.(\ref{four}) outperforms Eq.(\ref{three}).}\label{Fig:11}
\end{figure}

The Schmidt ansatz is somewhat restrictive since the algorithm does not cover the entire six-qubit space. Therefore, it is possible that with the NN ansatz as in Eq.(\ref{ansatz}), the ML algorithm may find other new and non-trivial codes that outperform the repetition codes. Next, we explore this possibility. Using an NN ansatz with 6 input neurons, 6 hidden layers, and a width of 6 units, we ran the same GA and, to our surprise, found several non-trivial codes that outperformed the repetition codes. These codes are,
\small
\begin{equation}
\begin{aligned}
\ket{\chi(I)}\equiv&(\ket{zzz}+\ket{\bar{z}\bar{z}\bar{z}})\ket{zzz}+(\ket{z\bar{z}z}-\ket{\bar{z}z\bar{z}})\ket{zz\bar{z}}+\\&(\ket{zzz}-\ket{\bar{z}\bar{z}\bar{z}})\ket{z\bar{z}z}+(\ket{z\bar{z}z}+\ket{\bar{z}z\bar{z}})\ket{z\bar{z}\bar{z}};\\
\ket{\chi(II)}\equiv&(\ket{xxx}+\ket{\bar{x}\bar{x}\bar{x}})\ket{xxx}+(\ket{x\bar{x}x}-\ket{\bar{x}x\bar{x}})\ket{xx\bar{x}}+\\&(\ket{xxx}-\ket{\bar{x}\bar{x}\bar{x}})\ket{x\bar{x}x}+(\ket{x\bar{x}x}+\ket{\bar{x}x\bar{x}})\ket{x\bar{x}\bar{x}};\\
\ket{\chi(III)}\equiv&(\ket{yyy}+\ket{\bar{y}\bar{y}\bar{y}})\ket{yyy}+(\ket{y\bar{y}y}-\ket{\bar{y}y\bar{y}})\ket{yy\bar{y}}+\\&(\ket{yyy}-\ket{\bar{y}\bar{y}\bar{y}})\ket{y\bar{y}y}+(\ket{y\bar{y}y}+\ket{\bar{y}y\bar{y}})\ket{y\bar{y}\bar{y}}; 
\end{aligned}\label{four}
\end{equation}
\normalsize
The regions in the three-parameter space where these codes exhibit super-additivity are shown in Fig.\ref{Fig:10}. As seen from the figure, the regions are very similar to those seen in Fig.\ref{Fig:6}. This is interesting since Eq.(\ref{four}) and Eq.(\ref{Quantumcodes3}) (by padding it for the three-shot case by using dummy qubits) are not equivalent. There are also points where the Eq.(\ref{four}) outperforms the repetition codes (for instance at $p_1=0.0,p_2=0.002, p_3=0.382$). Other such points can be seen in Fig.\ref{Fig:11}, where we have shown points in the three-parameter space where either of the two sets Eq.(\ref{three}), Eq.(\ref{four}) display super-additivity. The points are color-coded according to the best performing codes among the two sets. We have further calculated the expression for the three-shot capacity of these codes.

\begin{widetext}
\begin{equation}\label{q34}
    \begin{aligned}
Q^{(3)}(\ket{\chi(I)},\Lambda_{\vec{p}})&=\frac{1}{3}\Bigg(2p_1^2p_2\log\left(2p_1^2p_2\right)+2p_1p_2^2\log\left(2p_1p_2^2\right)+2p_3^2p_0\log\left(2p_3^2p_0\right)+2p_3p_0^2\log\left(2p_3p_0^2\right)+\\
&2p_1p_2p_3\log\left(2p_1p_2p_3\right)+2p_1p_2p_0\log\left(2p_1p_2p_0\right)+2p_1p_3p_0\log\left(2p_1p_3p_0\right)+2p_2p_3p_0\log\left(2p_2p_3p_0\right)+\\
&p_1\left(p_1^2+p_2^2\right)\log\left(p_1\left(p_1^2+p_2^2\right)\right)+p_1\left(p_3^2+p_0^2\right)\log\left(p_1\left(p_3^2+p_0^2\right)\right)+p_2\left(p_3^2+p_0^2\right)\log\left(p_2\left(p_3^2+p_0^2\right)\right)+\\
&p_3\left(p_3^2+p_0^2\right)\log\left(p_3\left(p_3^2+p_0^2\right)\right)+p_2\left(p_1^2+p_2^2\right)\log\left(p_2\left(p_1^2+p_2^2\right)\right)+p_3\left(p_1^2+p_2^2\right)\log\left(p_3\left(p_1^2+p_2^2\right)\right)+\\
&p_0\left(p_3^2+p_0^2\right)\log\left(p_0\left(p_3^2+p_0^2\right)\right)+\left(p_1^2+p_2^2\right)p_0\log\left(\left(p_1^2+p_2^2\right)p_0\right)+\\&2p_1\left(p_1p_3+p_2p_0\right)\log\left(p_1\left(p_1p_3+p_2p_0\right)\right)+2p_1\left(p_1p_0+p_2p_3\right)\log\left(p_1\left(p_1p_0+p_2p_3\right)\right)+\\&2p_2\left(p_1p_3+p_2p_0\right)\log\left(p_2\left(p_1p_3+p_2p_0\right)\right)+2p_2\left(p_1p_0+p_2p_3\right)\log\left(p_2\left(p_1p_0+p_2p_3\right)\right)+\\&2p_3\left(p_1p_3+p_2p_0\right)\log\left(p_3\left(p_1p_3+p_2p_0\right)\right)+2p_3\left(p_1p_0+p_2p_3\right)\log\left(p_3\left(p_1p_0+p_2p_3\right)\right)+\\
&2p_0\left(p_1p_3+p_2p_0\right)\log\left(p_0\left(p_1p_3+p_2p_0\right)\right)+2p_0\left(p_1p_0+p_2p_3\right)\log\left(p_0\left(p_1p_0+p_2p_3\right)\right)\\&-2\left(p_1+p_2\right)\left(p_0+p_3\right)\log\left(\frac{1}{2}\left(p_1+p_2\right)\left(p_0+p_3\right)\right)
\\&-\left(\left(p_1+p_2\right)^2+\left(p_0+p_3\right)^2\right)\log\left(\frac{\left(p_1+p_2\right)^2+\left(p_0+p_3\right)^2}{4})\right)\Bigg)
\end{aligned}
\end{equation}
\end{widetext}
\normalsize
As earlier, symmetry of the quantum codes and that of the Pauli channels can be used to show that,
\small
\begin{align*}
    Q^{(3)}(\ket{\chi(II)},\Lambda(p_1,p_2,p_3))&=Q^{(3)}(\ket{\chi(I)},\Lambda(p_3,p_2,p_1))\\
    Q^{(3)}(\ket{\chi(III)},\Lambda(p_1,p_2,p_3))&=Q^{(3)}(\ket{\chi(I)},\Lambda(p_1,p_3,p_2))
\end{align*}
\normalsize

Numerically, the maximal gap attained by these quantum codes is approximately $ 0.00681535$; when $\vec{p}=(p_1,p_2,p_3)=(0.00824609, 0.220845, 0.0277404)$ (and other permutations of $p_1$, $p_2$ and $p_3$).

\section{Conclusion}\label{secV}

We showed that a neural network representation of quantum states coupled with a simple evolutionary strategy could find quantum codes with high coherent information for the class of Pauli Channels. We considered the three-parameter family of Pauli channels and found quantum channels that display super-additivity of two-shot coherent information. We further characterized the quantum codes that the algorithm learned and surprisingly found that the algorithm only learned three distinct quantum codes. With the exact form of the quantum codes at hand, we calculated the exact form of the two-shot coherent information with these codes and further mapped regions of super-additivity in the three-parameter space, thus obtaining the structure as shown in Fig.\ref{Fig:6}. Our analysis showed that certain Pauli channels show a large gap between single-shot quantum capacity and two-shot quantum capacity; this high gap of around $0.01$ is much larger than what is seen in other well-studied classes of quantum channels, and therefore, we think these Pauli channels will be of use in experimental studies of quantum capacities. 

For better structural visualization, we investigated the super-additivity property under the constant noise of one of the parameters ($p_3$), effectively obtaining two-parameter families, plotted in Fig. \ref{Fig:7}, that display super-additivity of two-shot coherent information. Furthermore, these studies perceived the existence of Pauli channels using which no quantum information can be communicated in a single-use, but through two uses, communication is viable, a prominent example of `$0+0>0$'. The channels having the above feature are plotted in Fig. \ref{Fig:8}.    

We also applied the same approach to the three channel-use case of Pauli channels. With a Schmidt ansatz the algorithm found three distinct quantum codes, which are variants of the famous repetition code. The super-additive region obtained by using these codes is shown in Fig. \ref{Fig:9}. Our inspection is quite consistent with the studies of the Depolarizing channels, which shows super-additivity under three channel-uses but not under two uses of the channel; in fact, this seems to remain true for other one-parameter quantum channels which are `close', in terms of the noise structure, to the depolarising channel as can be seen from Fig.\ref{Fig:6}. 

The true power of machine learning approaches lies in finding non-trivial solutions to complex problems. Recognizing the possibility of such non-trivial quantum codes with high super-additivity, we used the NN ansatz in Eq.(\ref{ansatz}) to search through the 6 qubit Hilbert space, eventually discovering the set Eq.(\ref{four}). In fact, there are Pauli channels where this code outperforms the repetition codes.

The results shed light on the behavior and properties of the class of Pauli channels; this would be of much use in the context of near-term quantum computing, considering the usefulness of Pauli channels as a practical model for real quantum noise. Moreover, as discussed before, the high super-additivity displayed by some of the Pauli channels makes them an ideal candidate for experimental studies. The success of the GA based optimization technique with Neural Networks in finding new and non-trivial quantum codes is yet another instance of the power of machine learning techniques in aiding fundamental research and, more importantly, quantum information research.

 Note that these calculations become prohibitive as the number of channel uses increases; therefore, an interesting problem would be finding efficiently computable fitness measures that approximate coherent information. All our calculations involved basic techniques from the respective fields, and their success suggests that it is worthwhile to implement advanced GA-based optimization techniques and neural network representations to find more efficient and computationally feasible schemes. It would also be interesting to use these techniques to find generalizations of the codes in Eq.(\ref{Quantumcodes3}) and Eq.(\ref{four}) to higher channel uses, eventually finding better bounds to the quantum capacity of these channels.

{\bf Acknowledgment:} GLS is thankful for the DST-INSPIRE fellowship. MA and MB acknowledge funding from the National Mission in Interdisciplinary Cyber-Physical systems from the Department of Science and Technology through the I-HUB Quantum Technology Foundation (Grant no: I-HUB/PDF/2021-22/008). MB acknowledges support through the research grant of INSPIRE Faculty fellowship from the Department of Science and Technology, Government of India and the start-up research grant from SERB, Department of Science and Technology (Grant no: SRG/2021/000267). The authors thank the use of the {\it Padmanabha} computational cluster made available through the center for High Performance Computation at IISER-TVM.

\onecolumngrid
\appendix

\section{Analysis of learning performance}\label{Appendix}
In this section, we present the numerical performance of the optimization scheme used and show empirical evidence, which suggests that the NN-ansatz coupled with the presented GA algorithm is a promising approach to find quantum codes with high coherent information. In this section, we ask three main questions: (1) Is a NN-ansatz needed in the first place to find good quantum codes? (2) How does GA's learning performance change as the NN's size is varied? (3) How does the employed GA perform in comparison to the well-studied case of PSO?
\par
As we will soon elaborate on, we found through simulations that (1) A NN ansatz is not necessary for finding good quantum codes in the two-shot case. On the other hand, in the three-shot case, only the algorithms trained with a NN ansatz could find good quantum codes that showed super-additivity. This suggests that {\it a priori}, a NN ansatz is a good starting point for exploring the super-additivity of quantum channels. (2) Given fixed parameters, GA showed improved learning performance as the depth of the NN increased. We also found a positive correlation between learning performance and population size. This contrasts with a simple PSO variant, which performed worse as the depth of the NN increased. (3) By comparing the performance of GA and the PSO variant in test cases, we found that their performances are comparable. In particular, the employed PSO algorithm always performed superior to the GA when a RAW ansatz is used. RAW here refers to parameterize a quantum code explicitly by its coefficients in the computational basis. In sharp contrast, GA outperformed the PSO algorithm when the NN ansatz was used.
\par
Before elaborating on these points, we briefly introduce the PSO algorithm used in this paper.

\subsection{Particle Swarm Optimisation}

Particle Swarm Optimisation (PSO) is yet another derivative-free meta-heuristic search method that is extensively used in global optimization problems. A variant of PSO for the quantum codes search was extensively used in Ref.\cite{Bausch2020}.
\par
PSO works in the following mechanism: Initially, a set of candidate solutions (here called particles) are randomly generated in the optimization landscape. Each particle will be indexed by `i' and equipped with a position vector ($x_i$) encoding the candidate solution and a velocity vector ($v_i$) encoding the particle's direction of motion in the optimization landscape. Each particle also `remembers' the best position it has been in the landscape and the best position achieved (considering all the particles) since the start of the optimization. We also set an upper and lower-bound to the velocity components that can be achieved by each of the particles.
\par
In below, a snippet of the PSO scheme is given. The update equation is the same as given in Ref.\cite{Bausch2020}.

\begin{lstlisting}[language=Python,mathescape=True]
N_GEN #Number of generations
N_Particles #Number of particles
$\alpha$ # Intertia Parameter
$\beta$ # Self-Interaction Parameter
$\gamma$ # Mutual-Interaction Parameter
$x^B_i$ # Best position in particle-i's history
$x^G$ # Best position found in the entire history.
$v_{min}$, $v_{max}$ # Min and Max of velocity. 

Initialize all particles randomly

Repeat N_GEN times:
   
   For each particle-i:
      For each position/velocity component-k:
   
       $\beta_r\leftarrow \beta$*random(0,1)
       $\gamma_r\leftarrow \gamma$*random(0,1)
            
       $v_i(k) \leftarrow \alpha v_i(k) + \beta_r (x^{B}_i(k)-x_i(k)) +\gamma_r (x^{G}(k)-x_i(k)) $
      
       if $v_i(k)>v_{max}$:
         $v_i(k)\leftarrow v_{max}$
       if $v_i(k)<v_{min}$:
         $v_i(k) \leftarrow v_{min}$
        
       $x_i(k)\leftarrow x_i(k)+v_i(k)$
      end loop
      
      if Fitness($x_i$)>Fitness($x^{B}_i$):
         $x^{B}_i\leftarrow x_i$
         
   end loop
   
   $x^{G}\leftarrow \text{arg}\max_i [\text{Fitness}(x_i)] $
end loop

return Fitness($x^G$)
   
\end{lstlisting}

The PSO algorithm described above was implemented in Python using the DEAP package\footnote{See: https://deap.readthedocs.io/en/master/examples/pso\_basic.html for an implementation of the above algorithm}.

Now we move to the analysis of the learning performance.

\subsection{Learning Performance}
\subsubsection{Two uses of Pauli Channel}
We first study the problem of finding optimal quantum codes with high 2-shot coherent information. Purified quantum codes then belong to a 4-qubit Hilbert space. We first investigate if employing a NN ansatz is helpful for this problem. For this, we compare the learning performance with GA and PSO with a RAW ansatz and NN ansatzes of varying depth. As mentioned before, in the RAW ansatz, we treat the coefficients of the state vector (in the computational basis) as the optimization parameters. Throughout this section, we fix the quantum channel to be a Pauli channel with parameters p1=0.003, p2=0.285, and p3=0.008. This point belongs to the region (in Fig.\ref{Fig:6}) where we found quantum codes that display super-additivity of coherent information. 
\par
The Genetic algorithm was executed with 100 individuals, which evolved for 300 generations. As described in the main text, a `2D cross-over' type recombination strategy was used with a cross-over probability of 0.5. A Gaussian mutation of mean 0.5 and standard deviation of 0.25 was used to mutate individuals. The mutation scheme is such that each individual has a chance of 0.2 for being picked for mutation. Once picked, each of its attributes (encodings) has a chance of 0.5 for undergoing a Gaussian addition.
\par
The PSO algorithm was executed with 100 particles which evolved for 300 generations. We also set $\alpha=\beta=\gamma=0.5$. Maximum and minimum velocity of each component was set to 1 and -1 respectively.
\par
\begin{figure}[t]
    \subfloat[GA]{%
      \includegraphics[width=0.45\textwidth]{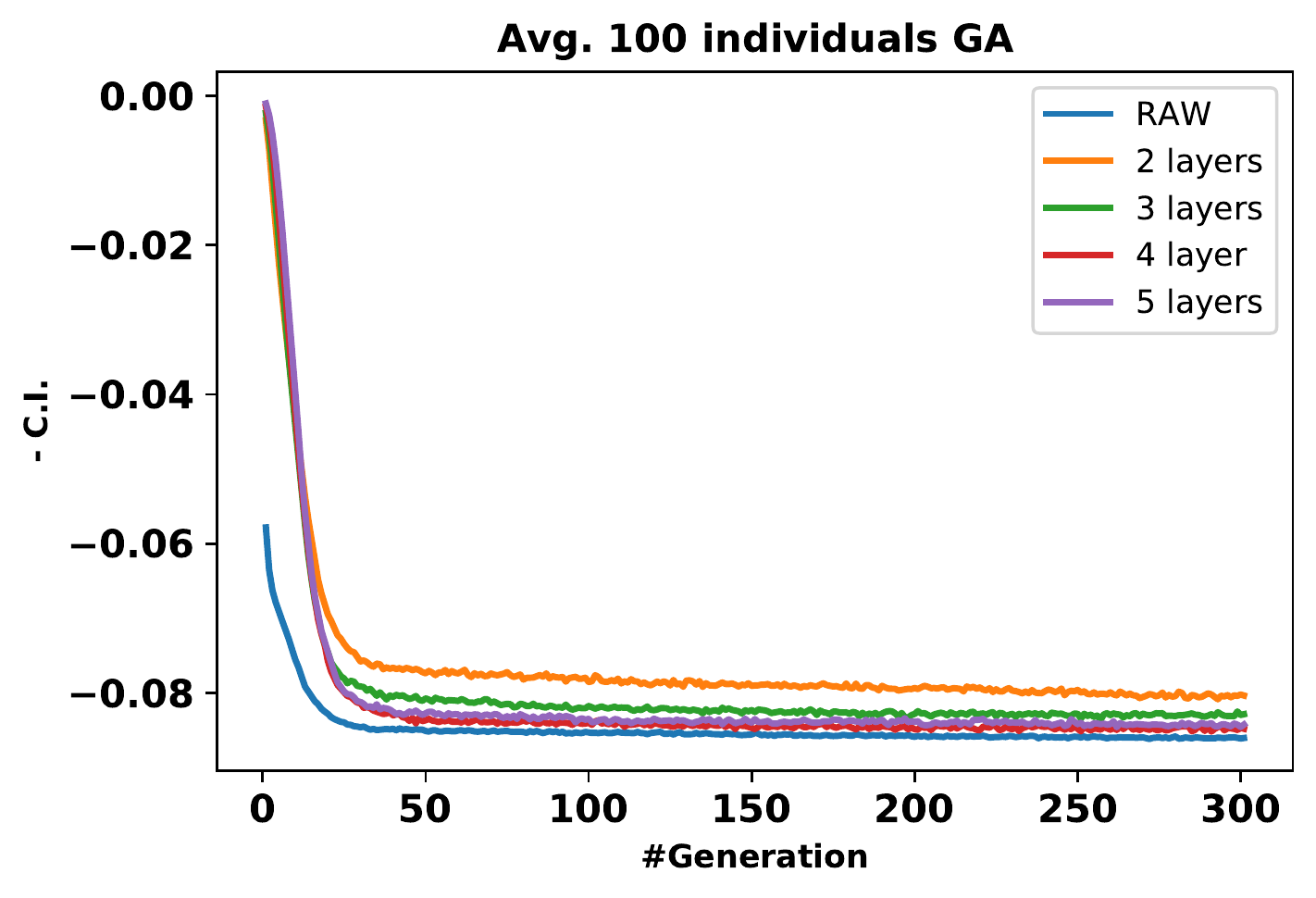}
    }
    \hfill
    \subfloat[PSO]{%
      \includegraphics[width=0.45\textwidth]{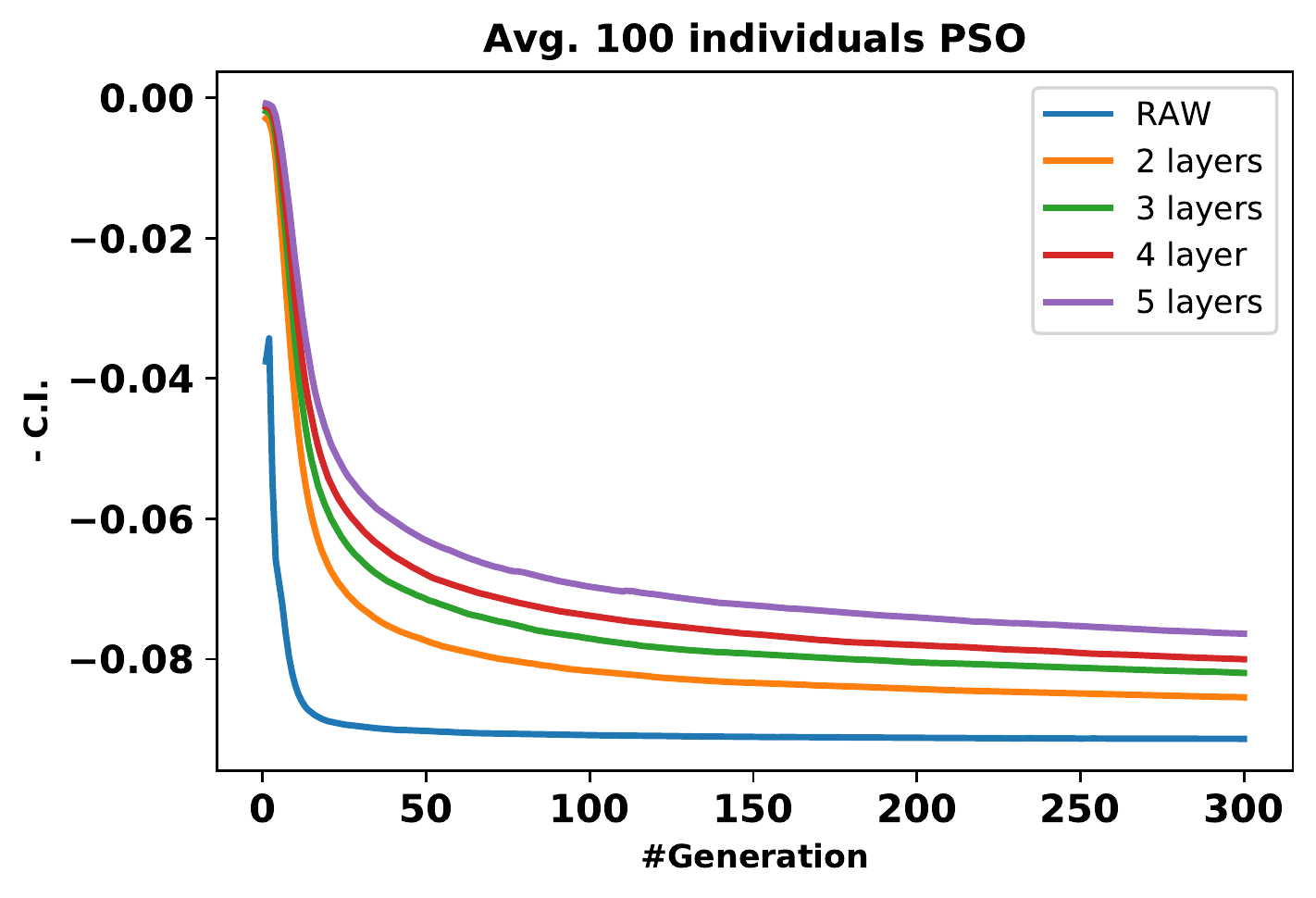}
    }
    \caption{(Color online) Learning curves for GA/PSO with NNs of increasing depth. A population of 100 individuals are evolved in both cases. RAW represent the case when the NN ansatz is not used and the coefficients of the wavefunction is treated as the optimisation variables. Each learning curve shows the Average Fitness of the population averaged over 100 learning trajectories. (a) Going down the y-axis near the asymptote, the lines are in the order '2 layers', '3 layers', '5 layers', '4 layers', 'RAW'. (b) (b) Top curve corresponds to  `5 layers' followed by `4 layers', `3 layers', `2 layers' and finally `RAW'.}
    \label{Fig:s1}
  \end{figure}

We compared the learning performance of GA and PSO with the RAW ansatz and Deep NN ansatzes of depths 5,4,3, and 2 layers. As described in the main text, these are fully connected NNs of a width of four units each. They are chosen to have a cosine activation function in the first layer and `tanh' activation function in all subsequent layers. The learning performance of these algorithms with varying NN depths is shown in Fig.\ref{Fig:s1}. As inferred from the figure, the population's average fitness consistently improves as the optimization proceeds in time. The first observation is that the RAW ansatz performs superior to an NN ansatz in both cases. There is also a significant difference between the performance of GA and PSO. As inferred from Fig.\ref{Fig:s1}, the performance of GA increases as the NN depth increases. This is in sharp contrast to the case of PSO, where the performance is observed to degrade as the NN's depth increases. 
\par
We also note that with a RAW ansatz, PSO performed better than GA, while with the NN ansatz, GA performed superior to PSO. This is a trend we observed throughout the analysis, as shall be mentioned in the coming sections. This does not, of course, mean that GAs are superior to PSO when equipped with an NN ansatz. For one, the employed PSO variant and GA are the simplest among a whole class of advanced PSO and GA variants. Moreover, even for the algorithms considered, the evidence provided is not decisive since the performance may change as the meta parameters are changed. So, at best, all we can conclude is that in general, PSO and GA are comparable for the optimisation task considered. We will return to this point later.

\begin{figure}[t]
    \subfloat[GA]{%
      \includegraphics[width=0.45\textwidth]{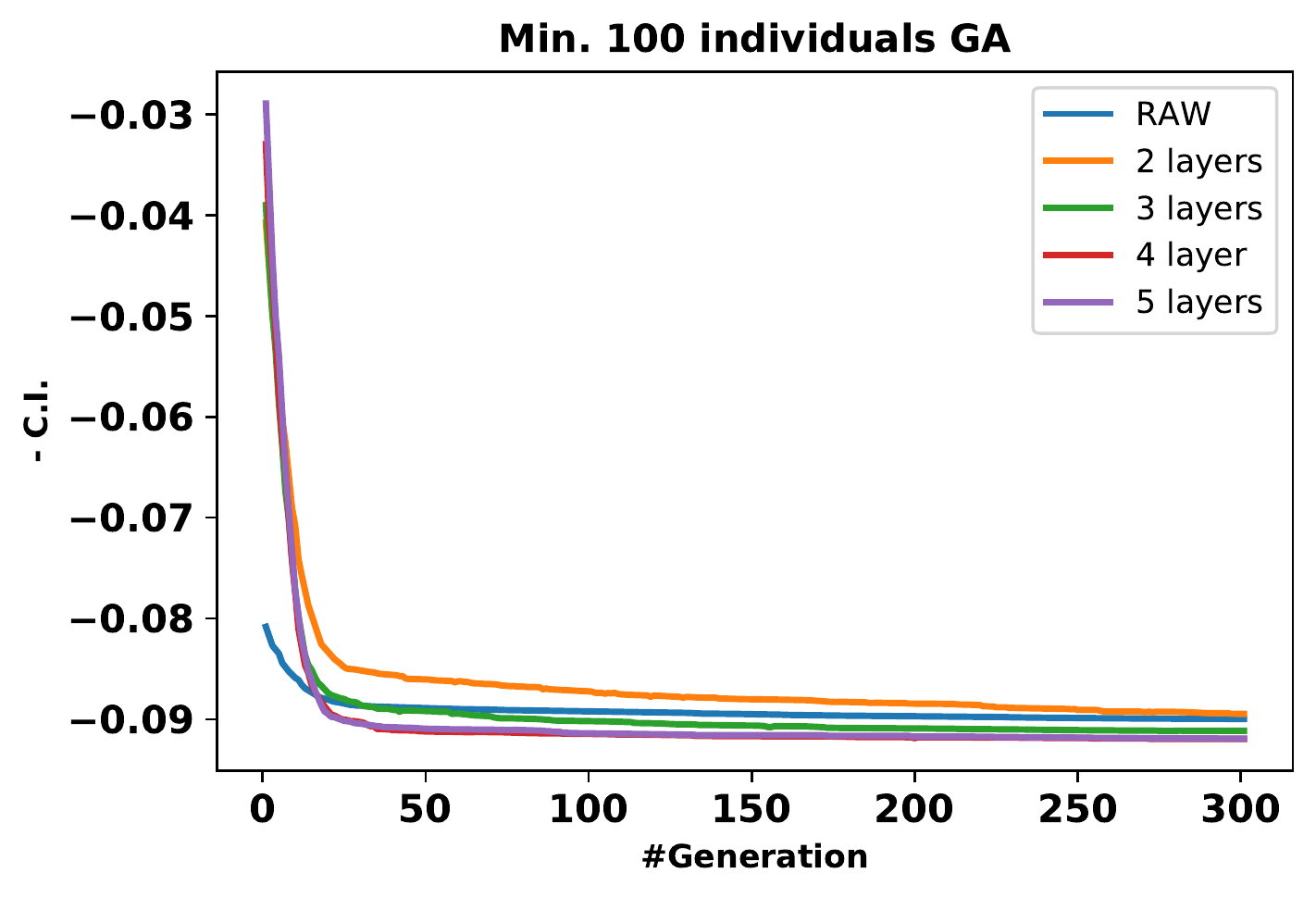}
    }
    \hfill
    \subfloat[PSO]{%
      \includegraphics[width=0.45\textwidth]{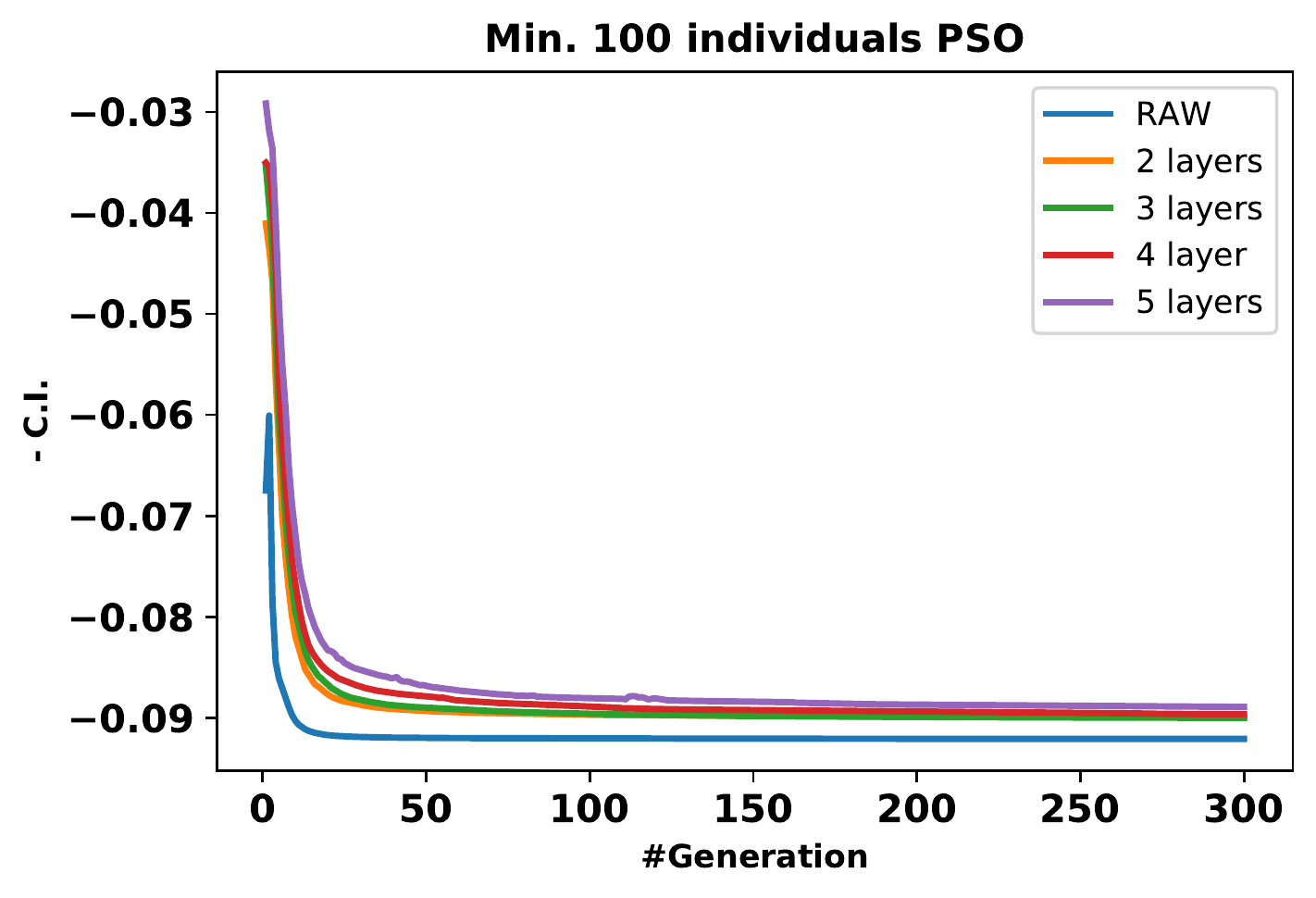}
    }
    \caption{(Color online) Learning curves for GA/PSO with NNs of increasing depth. A total of 100 individuals are evolved in each case. RAW represents the case when the NN ansatz is not used and the coefficients of the wavefunction is treated as the optimisation variables. Each learning curve shows the Best solution of a generation averaged over 100 learning trajectories. (a) Going down the y-axis near the asymptote, the lines are in the order '2 layers', 'RAW', '3 layers', '4 layers' and '5 layers'. (b) Top curve corresponds to  `5 layers' followed by `4 layers', `3 layers', `2 layers' and finally `RAW'.}
    \label{Fig:s2}
  \end{figure}

In the previous discussion, we looked at the population's average fitness as a measure of learning performance. Perhaps a more useful metric is the best solution found at each generation. The learning curves with this new metric are shown in Fig.\ref{Fig:s2}. As inferred from the figure, the main observations from the previous discussions follow over to this case as well.

\begin{figure}[h]
    \subfloat[Population size:50]{%
      \includegraphics[width=0.45\textwidth]{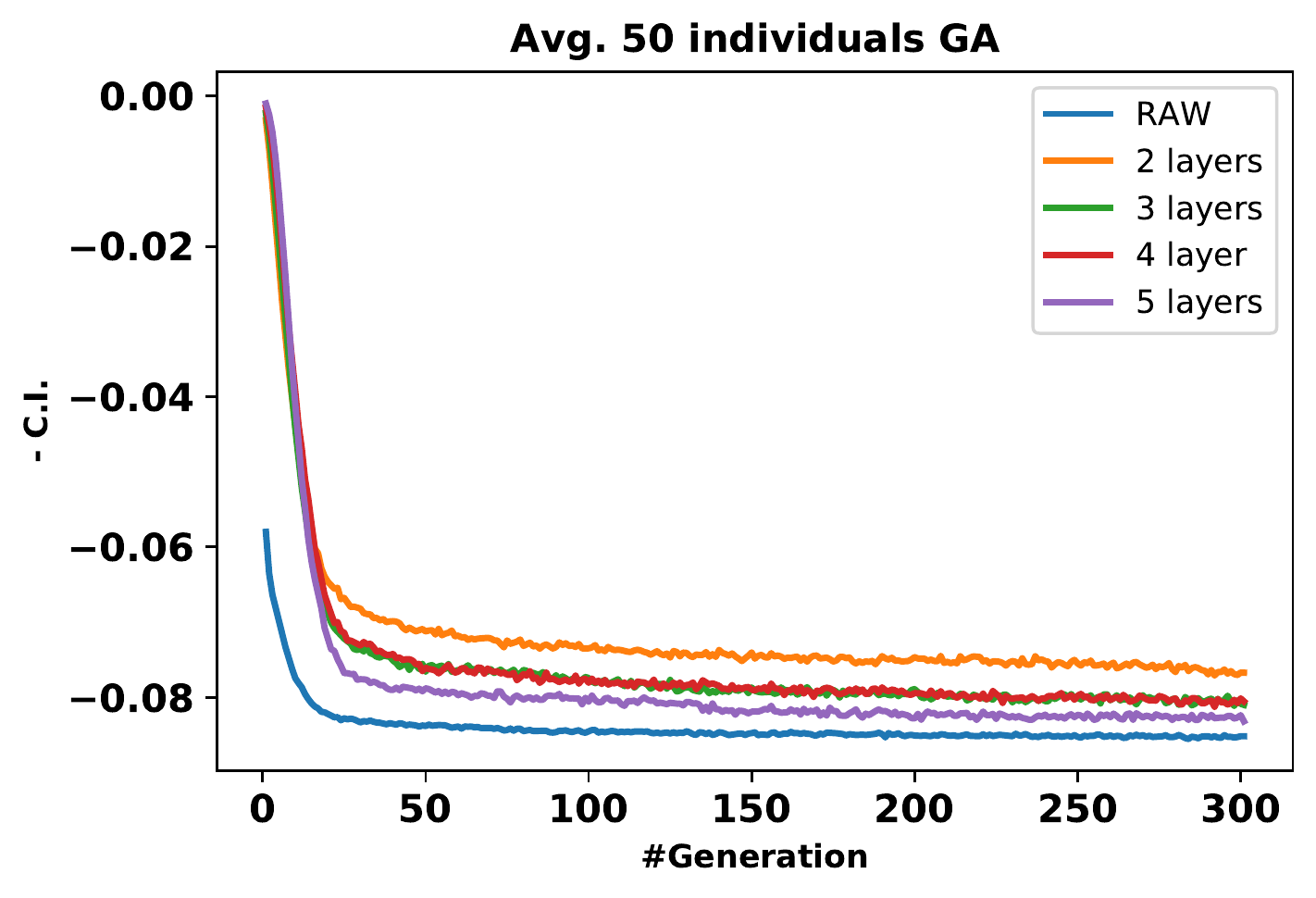}
    }
    \hfill
    \subfloat[Population size:300]{%
      \includegraphics[width=0.45\textwidth]{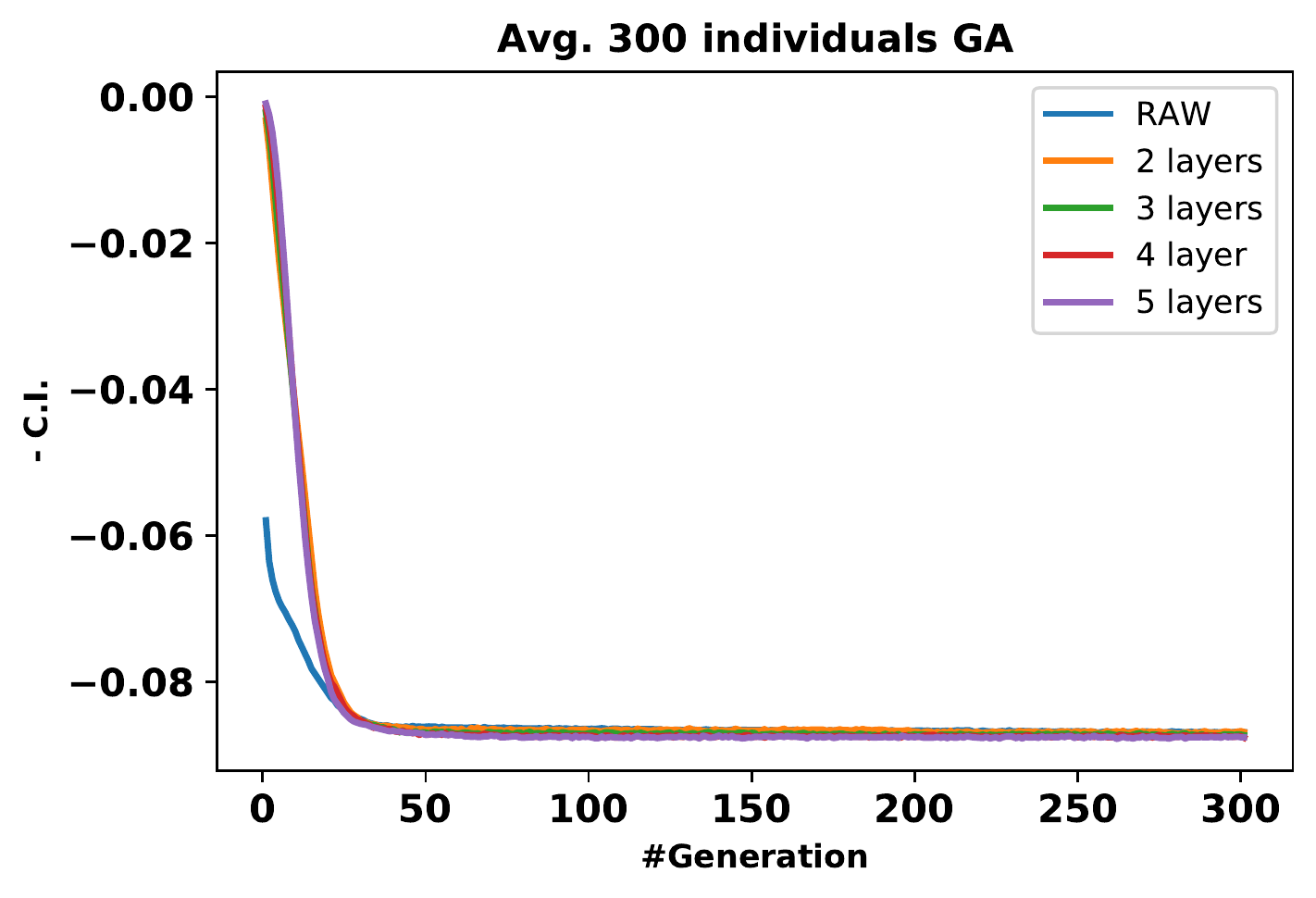}
    }
    \caption{(Color online) Learning curves for GA with NNs of increasing depth. Each learning curve shows the average fitness of the population averaged over 100 learning trajectories. (a) The initial population size is set to 50 individuals. The top curve corresponds to '2 layers', followed by '3 layers', '4 layers', '5 layers' and finally 'RAW' at the bottom. (b)The initial population size is set to 300 individuals. The high population size resulted in an insensitivity to NN depth and hence the learning curves are overlapped and indistinguishable.}
    \label{Fig:s3}
  \end{figure}

It is also interesting to see how the population size affects learning performance. For this, we re-did the previous calculations with a population size of 50 and 300 and found that in all these cases, the main observations made earlier carry over without fail. The corresponding learning curves for GA are shown in Fig.\ref{Fig:s3}. The learning curves for the PSO algorithm is relatively insensitive to different population sizes and are hence not shown. It is clear by comparing Fig.\ref{Fig:s1} to Fig.\ref{Fig:s3} that as the population size increases, the sensitivity of learning performance to the depth of the NN decreases to the point of saturation at a population of 300. This shows a trade-off between population size and depth of the NN as intuitively expected. 

\begin{figure}[h]
    \subfloat[GA]{%
      \includegraphics[width=0.45\textwidth]{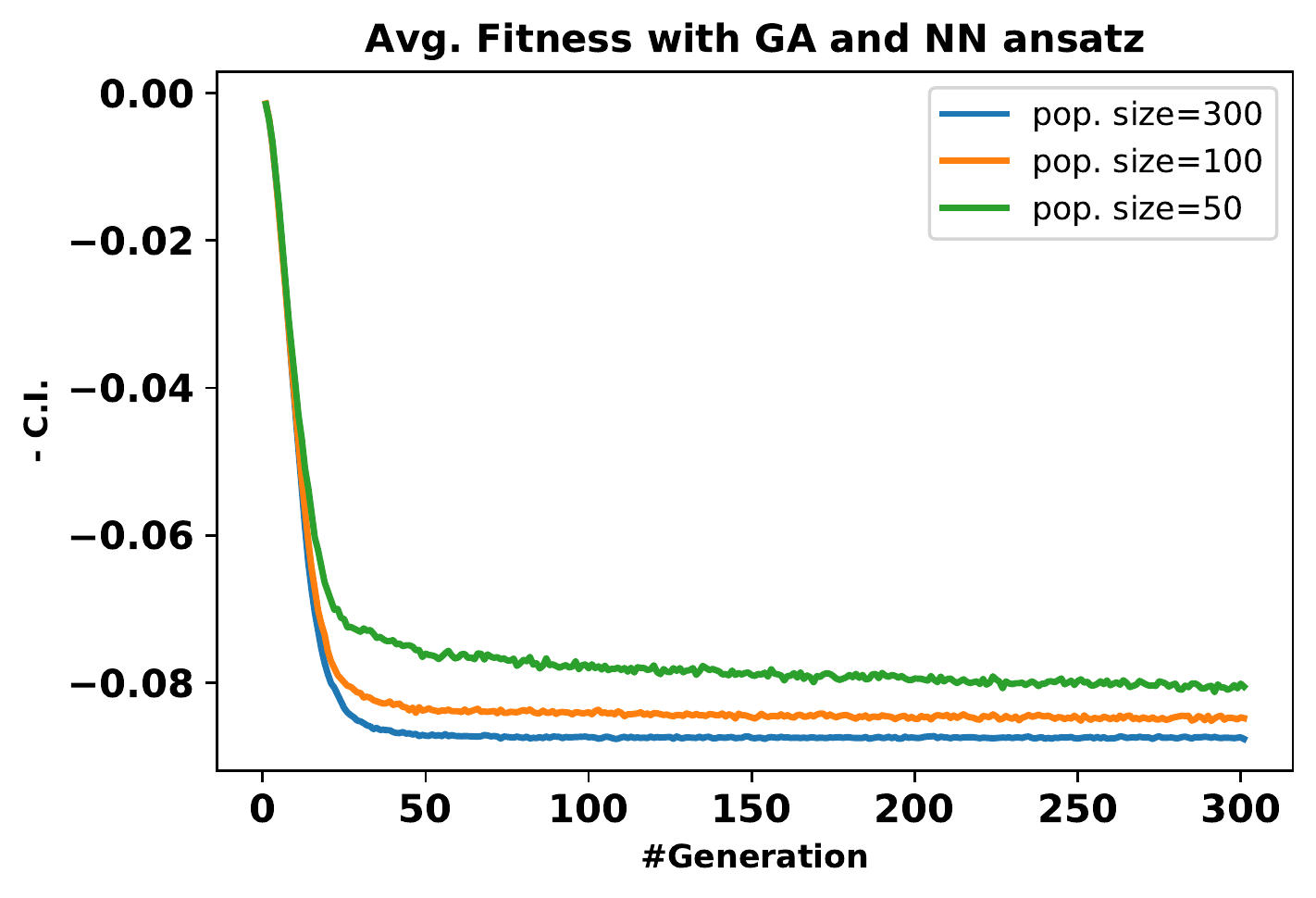}
    }
    \hfill
    \subfloat[PSO]{%
      \includegraphics[width=0.45\textwidth]{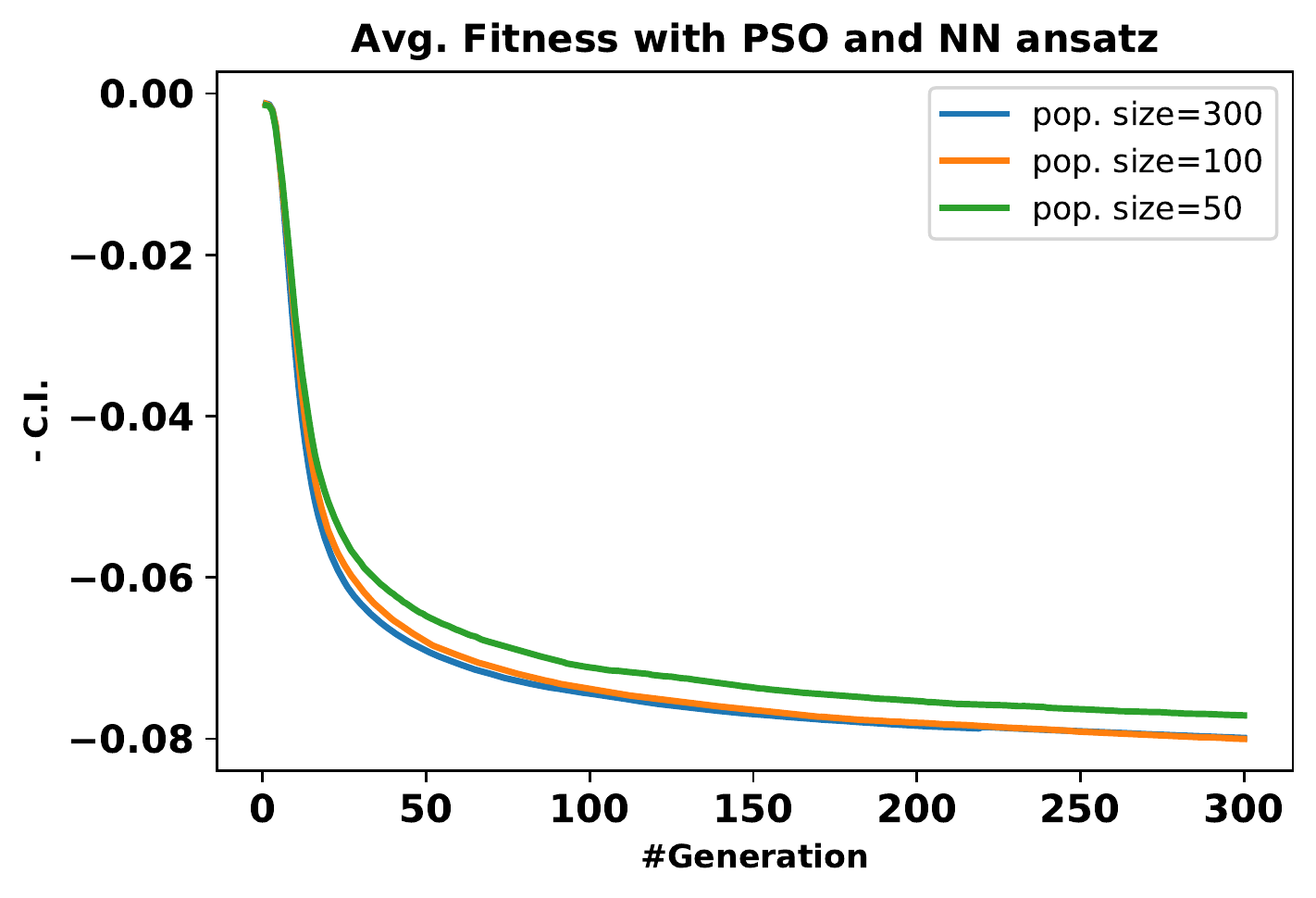}
    }
    \caption{(Color online) Learning curves for GA and PSO with increasing population size. A NN with a depth of 4 layers is used as the optimisation ansatz. Each learning curve shows the average fitness of the population averaged over 100 learning trajectories. (a) Population size of 300 (bottom curve) shows optimal performance, followed by population size of 100 (middle) and then by a size of 50 (top). (b) Going down the y-axis near the asymptote, the lines are in the order 'pop. size=50', 'pop. size=100', 'pop. size=300'.}
    \label{Fig:s4}
  \end{figure}

It would also be illuminating to see the dependence of learning performance on population size. This is shown in Fig.\ref{Fig:s4}, where the population dependence of GA/PSO was studied with a NN ansatz of depth 4. As intuitively expected, we observed a positive correlation between learning performance and population size. The trend is observed irrespective of the algorithm (GA/PSO), ansatz (RAW/NN), or learning metric (Avg./Min.) used. Therefore, only the learning performance of GA and PSO with a 4-layered NN ansatz and Average fitness metric is shown in Fig.\ref{Fig:s4}. We also point out an observed relative insensitivity of the employed PSO scheme to population size. 

  The preceding discussion focused on two-shot coherent information. A recurring observation we made is that in this case, the kind of NN ansatz we used is not of much use when compared with RAW ansatz. This is likely due to the small size of the 4-qubit Hilbert space. We also saw that in this case, the GA scheme employed improves performance as the depth of the NN is increased or as the population size is increased. We also compared GA with a simple PSO variant and showed that the performance of both algorithms is comparable. Interestingly, we saw a degradation of PSO's performance with increased depth of the NN. Similar to GA, the performance of PSO was seen to improve with increased population size; still, the sensitivity to the population size is weaker compared to the GA.
  
  \subsubsection{Three uses of Pauli Channel}
  
\begin{figure}[t!]
    \subfloat[GA]{%
      \includegraphics[width=0.45\textwidth]{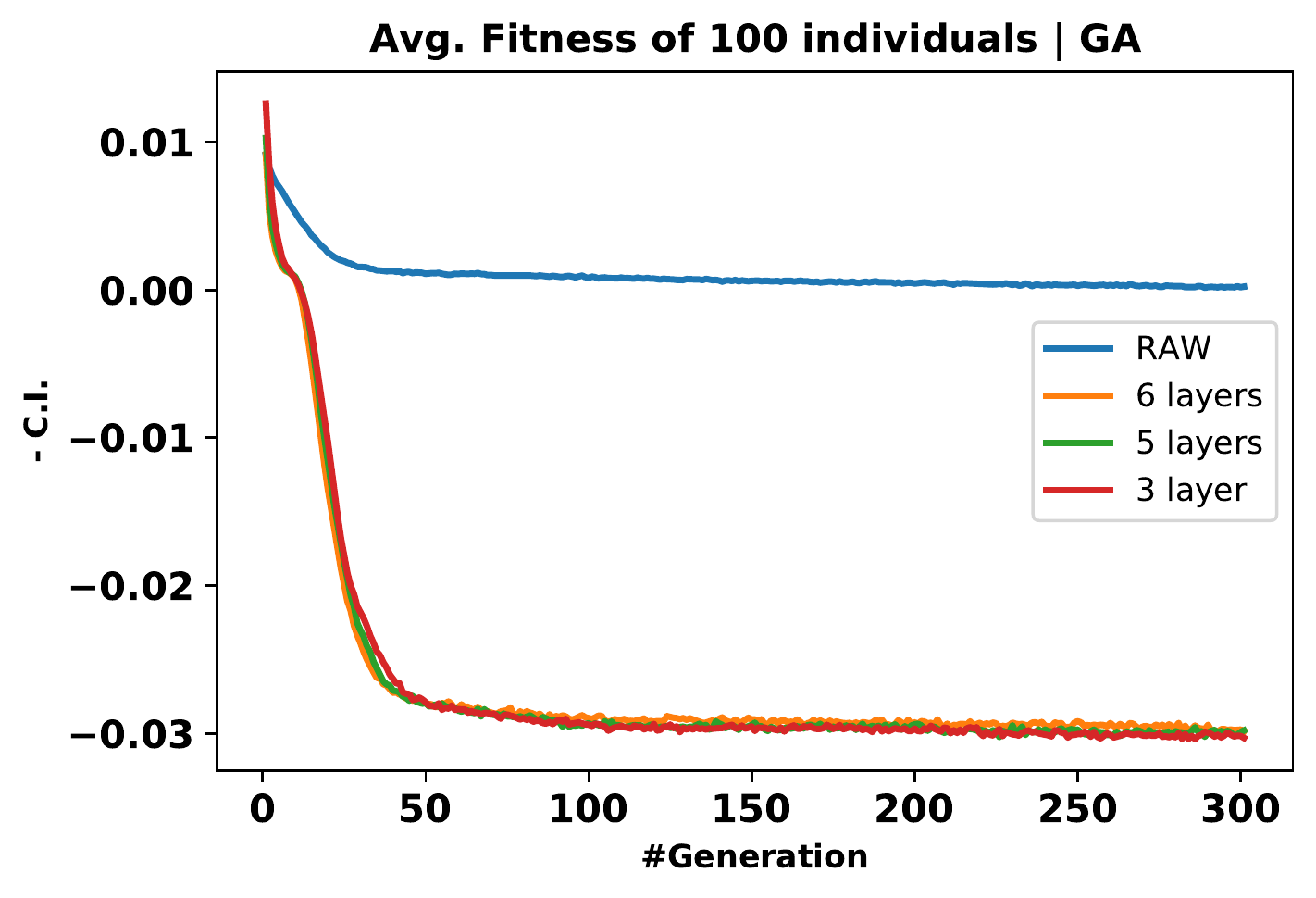}
    }
    \hfill
    \subfloat[PSO]{%
      \includegraphics[width=0.45\textwidth]{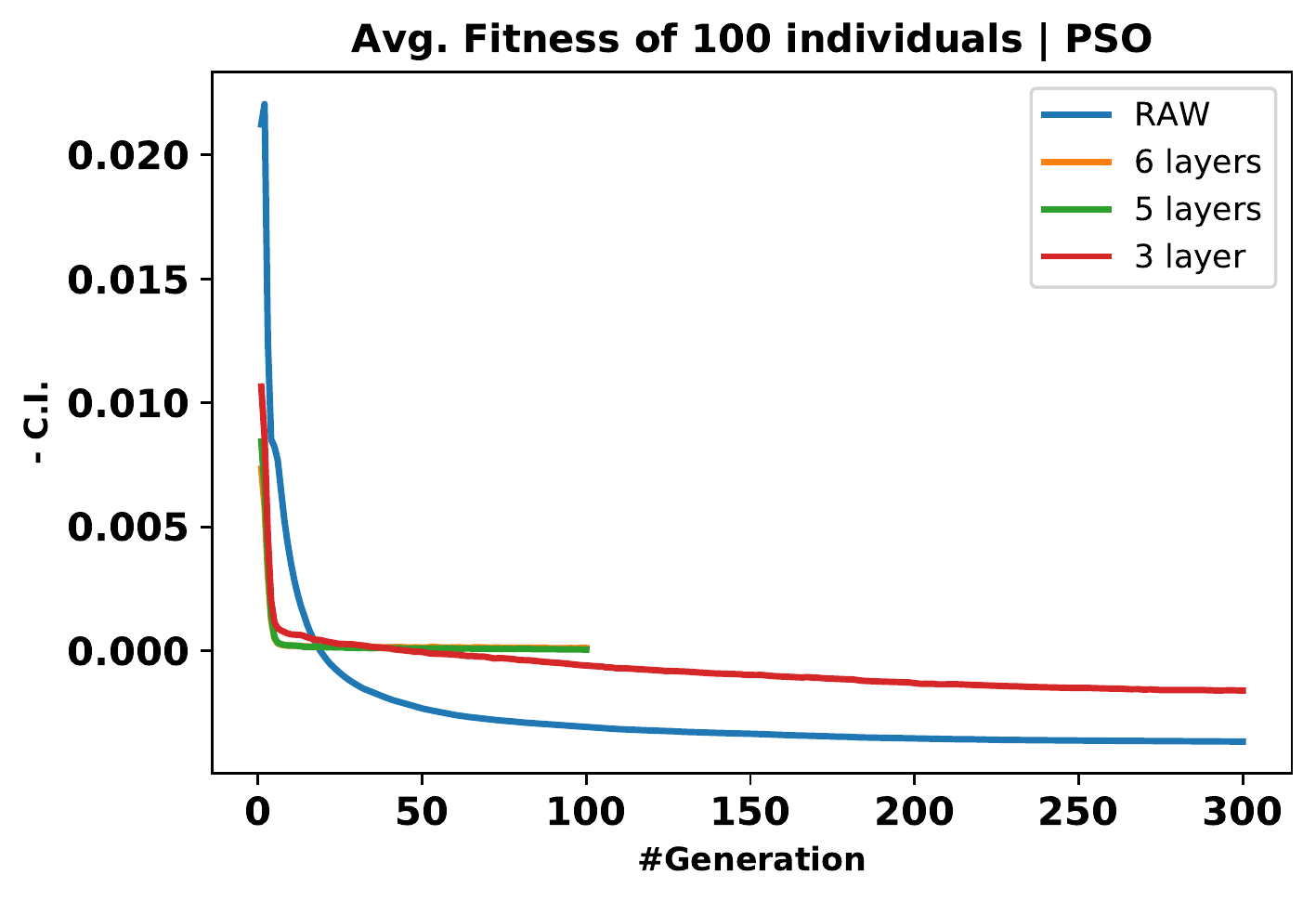}
    }
    \caption{(Color online) Learning curves for GA/PSO with NNs of increasing depth. RAW represent the case when the NN ansatz is not used and the coefficients of the wavefunction is treated as the optimisation variables. Each learning curve shows the Average Fitness of the population averaged over 100 learning trajectories. (a) The top curve corresponds to RAW ansatz. Learning curves corresponding to 6 layers, 5 layers and 3 layers are largely indisinguishable. (b) The top curve corresponds to '6 layers', followed by '5 layers', '3 layers' and finally the RAW ansatz.}
    \label{Fig:s5}
  \end{figure}

  Now we look at the 3-shot problem. The quantum codes now belong to a 6-qubit Hilbert space. We will consider the Pauli channel with parameters $p_1=0.006, p_2=0.022, p_3=0.247$ as a testbed to study the performance of the various schemes. This is the Pauli channel for which we found the quantum codes with the highest 3-shot super-additivity. We start with the important question: Can a NN ansatz outperform RAW optimization in this 6-qubit problem? To answer this, we first studied the learning dynamics of GA/PSO with NNs of varying depth and compared it with RAW optimization. The meta-parameters of GA and PSO are the same as used in the previous section. The resulting learning curves are shown in Fig.\ref{Fig:s5}, where RAW optimization is compared with three NNs with depths 6,5, and 3. The width of each NN is set to 6 units, and the activation functions are chosen as earlier.
  
  From Fig.\ref{Fig:s5}, it is clear that NN ansatz performs superior to the RAW optimization in the three-shot problem. We also note while the employed GA scheme coupled with a NN ansatz finds quantum codes of high coherent information, the PSO variant fails at this. In the case of RAW optimization, however, PSO performs slightly better than GA; yet no code of high coherent information was found. We also observe that the learning curves for GA are largely insensitive to the depth of the NN. The learning curves behave similarly with a `Best Solution' metric and are not shown here to avoid redundancy.
  
\begin{figure}[t!]
    \subfloat[GA]{%
      \includegraphics[width=0.45\textwidth]{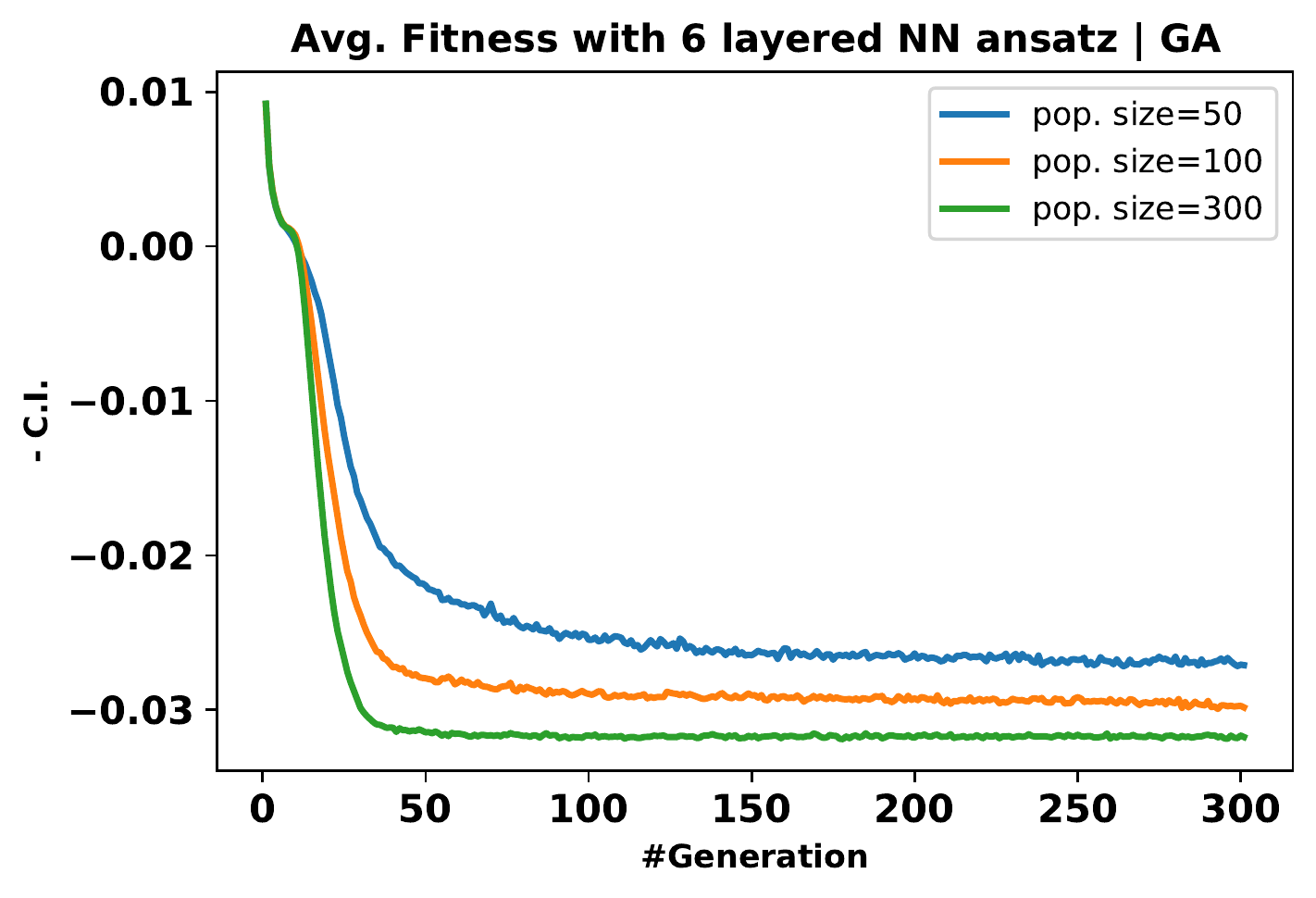}
    }
    \hfill
    \subfloat[PSO]{%
      \includegraphics[width=0.45\textwidth]{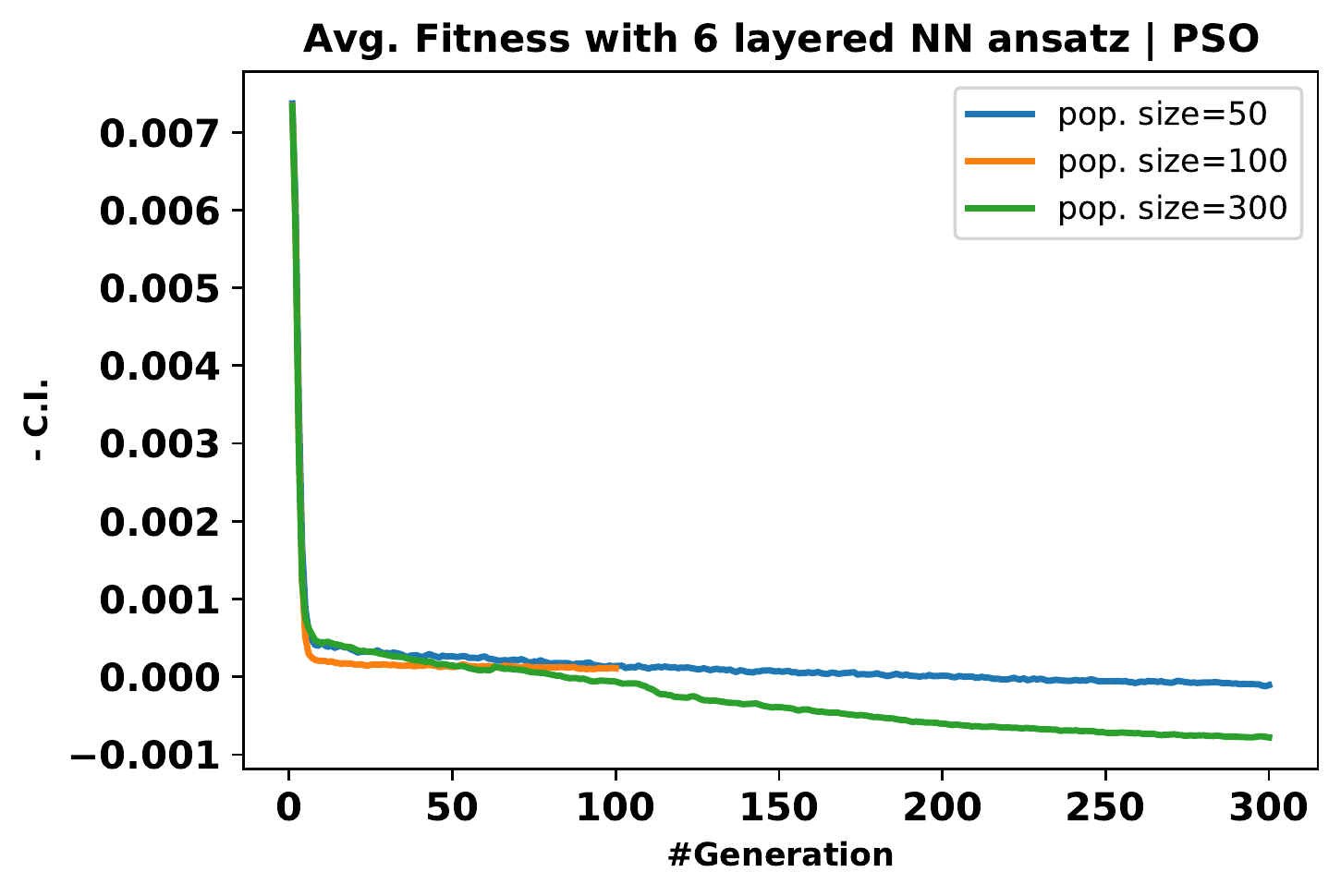}
    }
    \caption{(Color online) Learning curves for GA and PSO with increasing population size. A NN with a depth of 6 layers is used as the optimisation ansatz. Each learning curve shows the average fitness of the population averaged over 100 learning trajectories. (a) the top curve corresponds to a population size of 50, followed by population size of 100 and finally with a population size of 300 individuals. (b) The top curves, corresponding to population size of 50 and 100 are largely indistinguishable near the asymptote; these are followed by a green curve corresponding to a population size of 300.}
    \label{Fig:1s6}
  \end{figure}

  We now look at the effect of population size on learning performance. For this, we fix an NN ansatz of depth six and analyze the learning performance with populations of sizes 300,100, and 50. The results are shown in Fig.\ref{Fig:1s6}. As intuitively expected, performance improvement is observed as the population size is increased. As earlier, the employed PSO scheme is relatively insensitive to population size.
  
  The study revealed that a NN ansatz outperforms RAW optimization for finding quantum codes with high 3-shot coherent information. This displays the utility of machine learning approaches to finding good quantum codes in optimization problems, consistent with the results obtained in Ref.\cite{Bausch2020}.
  \par
   
  \begin{figure}[t!]
      \centering
      \includegraphics[width=0.45\textwidth]{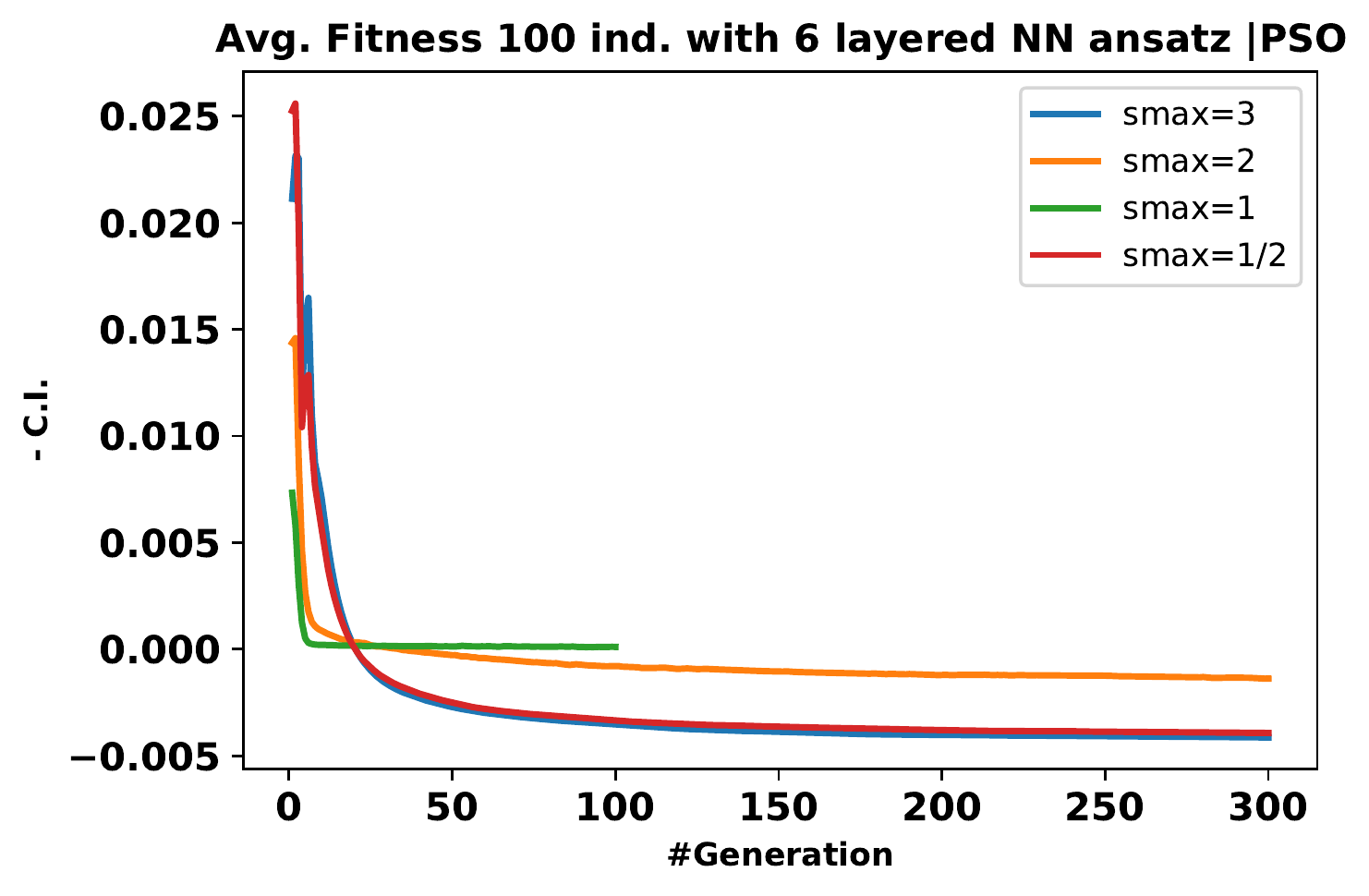}
      \caption{(Color online) Learning curves for PSO with different max. velocities. A 6-layered NN is used as the optimization ansatz. Each learning curve shows the average fitness of the population, averaged over 100 learning trajectories. The top curve corresponds to smax=1, followed by the curve with a maximum speed of 2. The curves corresponding to smax=3 and smax=1/2 are largely indistinguishable from each other and show maximum performance.}
      \label{Fig:1s7}
  \end{figure}
  
  Throughout the preceding discussion, we also observed that with the RAW ansatz, the employed PSO variant outperforms the GA. In contrast, we saw that when a NN ansatz is used, the GA was seen to outperform the PSO scheme significantly. We also noticed that the performance of GA improves with the depth of NN ansatz, while PSO displayed degradation with depth. However, these observations should be treated with caution as this does not imply that GAs outperform PSO schemes in all cases of interest. For one, we used the simplest variants of GAs and PSO schemes; things might significantly change when more advanced methods are used. Moreover, we have not explored the learning performance for all possible values of the associated meta-parameters. There might be different meta-parameter settings where PSO will perform comparable to, or even better than, GA with the NN ansatz. To illustrate the dependence on meta-parameters, we have evaluated the learning curves of PSO with varying max velocity. This is shown in Fig.\ref{Fig:1s7}. As observed from the figure, the learning performance depends (rather non-trivially) on the maximum allowed speed for each velocity component.

 Compiling all these observations suggests that a GA coupled with a NN ansatz significantly outperforms RAW optimization and is, therefore, a good optimization scheme for finding quantum codes of high coherent information. We also saw that, in the chosen meta-parameter settings, GA is more robust for use with an NN ansatz compared to the RAW ansatz. We also pointed out that this should not be taken for the superior performance GA schemes to PSO schemes. At best, we may conclude that for the problem at hand, the two methods are comparable with each other.

\twocolumngrid

\end{document}